\RequirePackage{fix-cm}
\documentclass[twocolumn,epjc3]{svjour3}
\smartqed  
\let\oldbibitem\bibitem
\def\bibitem{\vfill\oldbibitem}
\RequirePackage{amsmath,amssymb}
\RequirePackage{graphicx}
\RequirePackage{mathptmx}
\RequirePackage{flushend}
\RequirePackage[colorlinks,citecolor=blue,urlcolor=blue,linkcolor=blue]{hyperref}
\RequirePackage{siunitx}
\RequirePackage{booktabs}
\renewcommand{\eqref}[1]{eq.~(\ref{#1})}
\journalname{Eur. Phys. J. C}
\input{definitions}
\begin{document}
\title{A pulsed, mono-energetic and angular-selective UV photo-electron source for the commissioning of the KATRIN experiment}
\titlerunning{A photo-electron source for the commissioning of the KATRIN experiment}
\author{%
J.~Behrens\thanksref{e1,addr1,addr2now}
\and P.~C.-O.~Ranitzsch\thanksref{e2,addr1}
\and M.~Beck\thanksref{addr1,addr3}%
\and A.~Beglarian\thanksref{addr5}%
\and M.~Erhard\thanksref{addr2}%
\and S.~Groh\thanksref{addr2}%
\and V.~Hannen\thanksref{addr1}%
\and M.~Kraus\thanksref{addr2}%
\and H.-W.~Ortjohann\thanksref{addr1}%
\and O.~Rest\thanksref{addr1}%
\and K.~Schl{\"o}sser\thanksref{addr4}%
\and T.~Th{\"u}mmler\thanksref{addr4}%
\and K.~Valerius\thanksref{addr1,addr4}%
\and K.~Wierman\thanksref{addr6}%
\and J.~F.~Wilkerson\thanksref{addr6}%
\and D.~Winzen\thanksref{addr1}%
\and M.~Zacher\thanksref{addr1}%
\and C.~Weinheimer\thanksref{addr1}%
}
\institute{%
WWU M\"unster, Institut f\"ur Kernphysik, Wilhelm-Klemm-Str. 9, D-48149 M\"unster, Germany \label{addr1}%
\and IEKP, Karlsruhe Institute of Technology, Hermann-von-Helmholtz-Platz 1, D-76344 Eggenstein-Leopoldshafen, Germany \label{addr2}%
\and Institut f{\"u}r Physik, Johannes-Gutenberg Universit{\"a}t, D-55099 Mainz, Germany \label{addr3}%
\and IKP, Karlsruhe Institute of Technology, P.O. box 3640, D-76021 Karlsruhe, Germany \label{addr4}%
\and IPE, Karlsruhe Institute of Technology, Hermann-von-Helmholtz-Platz 1, D-76344 Eggenstein-Leopoldshafen, Germany \label{addr5}%
\and University of North Carolina, Department of Physics and Astronomy, Phillips Hall, CB 3255, Chapel Hill, NC 27599-3255, USA \label{addr6}%
}
\thankstext{e1}{e-mail: jan.behrens@kit.edu}
\thankstext{e2}{e-mail: philipp.ranitzsch@uni-muenster.de}
\thankstext{addr2now}{Now at [2].}
%
%
\date{Received: date / Accepted: date}
\maketitle
\begin{abstract}
The KATRIN experiment aims to determine the neutrino mass scale with a sensitivity of \SI{200}{meV/c^2} (\CL{90}) by a precision measurement of the shape of the tritium $\beta$-spectrum in the endpoint region.
The energy analysis of the decay electrons is achieved by a MAC-E filter spectrometer.
To determine the transmission properties of the KATRIN main spectrometer, a mono-energetic and angular-selective electron source has been developed.
In preparation for the second commissioning phase of the main spectrometer, a measurement phase was carried out at the KATRIN monitor spectrometer where the device was operated in a MAC-E filter setup for testing.
The results of these measurements are compared with simulations using the particle-tracking software ``Kassiopeia'', which was developed in the KATRIN collaboration over recent years.

%
\end{abstract}
%
%
\section{Introduction}
\label{intro}

The \textbf{KA}rlsruhe \textbf{T}ritium \textbf{N}eutrino experiment \textbf{KATRIN}~\cite{DesignReport2005} aims to measure an `effective mass' of the electron anti-neutrino, given by an incoherent sum over the mass eigenstates~\cite{NuMass2013}. It performs kinematic measurements of tritium $\beta$-decay to achieve a neutrino mass sensitivity down to $\SI{200}{meV/c^2}$ at \CL{90}, improving the results of the predecessor experiments in Mainz~\cite{NuMassMainz2005} and Troitsk~\cite{NuMassTroitsk2011} by one order of magnitude.
As the evolution of the neutrino mass results of these experiments showed, the study of systematic effects is of major importance: Underestimated or unknown ``energy loss'' processes caused too positive or even negative values for the square of the neutrino mass~\cite{NuMass2008}.
A detailed understanding of systematic uncertainties at the KATRIN experiment is crucial to achieve its target sensitivity.

\begin{figure*}[htb]
    \centering
    \includegraphics[width=.9\textwidth]{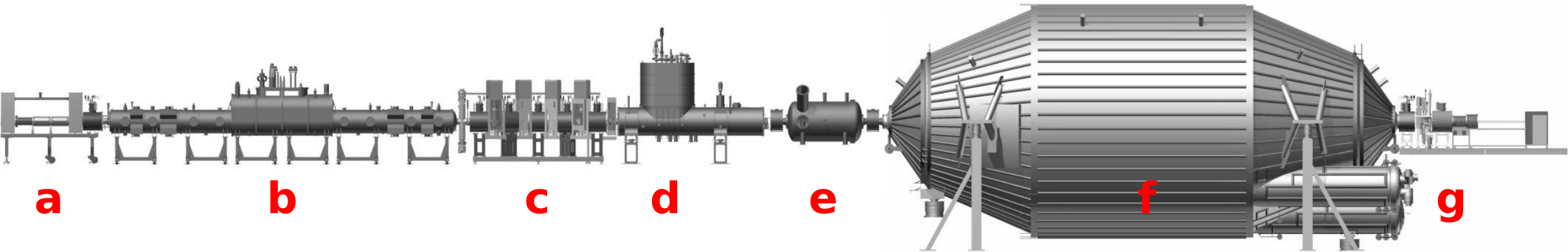}
    \caption{The beamline of the KATRIN experiment. The electrons are created via tritium $\beta$-decay inside the WGTS (b). The rear section (a) contains calibration tools to determine the source parameters and for commissioning of the setup. The decay electrons are guided through the DPS (c) and CPS (d), where the tritium flow is reduced by 14 orders of magnitude. The pre-spectrometer (e) rejects the low-energy part of the decay spectrum. The electron energy is determined by the main spectrometer (f), which follows the MAC-E filter principle. An integral measurement is performed by determining the electron rate at the FPD (g) at different filter energies of the main spectrometer.}
    \label{fig:katrin}
\end{figure*}

The outline of the KATRIN experiment is depicted in figure~\ref{fig:katrin}~\cite{DesignReport2005,NuMass2013}.
Molecular tritium is fed into the \SI{10}{m} long beam tube of the windowless gaseous tritium source (WGTS~\cite{WGTS2015}).
Superconducting magnets along the beam line create an adiabatic guiding field in a \SI{191}{T.cm^2} magnetic flux tube, and $\beta$-decay electrons emitted in forward direction propagate towards the spectrometer section.
The electrons then enter the transport and pumping section that reduces the tritium flow by a factor of \num{e14} in total~\cite{DPS2006}, using a combination of a differential pumping section (DPS~\cite{DPS2012}) with turbo-molecular pumps and a cryogenic pumping section (CPS~\cite{CPS2010}) where tritium is adsorbed by an argon frost layer.
The kinetic energy of the decay electrons is analyzed in a tandem of MAC-E filter\footnote{Magnetic Adiabatic Collimation with Electrostatic filter.} spectrometers~\cite{MACE1992,MACE1985,MACE1981}.
The main spectrometer achieves an energy resolution of \SI{0.93}{eV} at the tritium endpoint of $E_0(\text{T}_2) = \SI{18571.8(12)}{eV}$~\cite{NuMass2008,TritiumEndpoint2015} by a combination of an electrostatic retarding potential and a magnetic guiding field. Electrons with sufficient kinetic energy pass the retarding potential and are counted at the focal-plane detector (FPD~\cite{FPD2015}). An integral energy spectrum is measured by varying the filter energy close to the tritium endpoint. The effective neutrino mass is determined by fitting the convolution of the theoretical $\beta$-spectrum with the response function of the spectrometer to the data, taking into account important parameters such as the final states distribution and the energy loss spectrum and other systematic corrections~\cite{DesignReport2005,WGTS2012}.
The spectrometer high-voltage is monitored by a pair of precision high-voltage dividers~\cite{HVdivider2013,HVdivider2009} that support voltages up to \SI{35}{kV} and \SI{65}{kV}, respectively. An absolute voltage calibration is achieved by measuring the divider's output voltage with \si{ppm} precision using a digital voltmeter.
Additionally, the stability of the retarding potential is monitored continuously at the monitor spectrometer~\cite{MonSpec2014}. This MAC-E filter is connected to the main spectrometer high voltage system and measures natural conversion lines of \isotope{Kr}{83m}, where changes in the retarding potential are observed as shifts in the measured line position.

A precise knowledge of the transmission properties of the KATRIN main spectrometer is crucial to limit systematic uncertainties and reach the desired neutrino mass sensitivity. The transmission properties are affected by inhomogeneities of the electromagnetic fields in the main spectrometer. In addition to simulations, dedicated measurements are necessary to determine the spectrometer transmission function over the complete magnetic flux tube. Such measurements require a mono-energetic and angular-selective electron source, which we present in this work. A pulsed electron beam allows us to access additional information from the electron time-of-flight (ToF)~\cite{ToF2013}.

This article is structured as follows:
Section~\ref{design} discusses the revised technical design of the photoelectron source that was developed at WWU M{\"u}nster over the recent years~\cite{Egun2009,Egun2011,Egun2014}. The design underwent many improvements for the second commissioning phase of the KATRIN main spectrometer.
In section~\ref{measurements} we show results from test measurements at the KATRIN monitor spectrometer. We determine important source characteristics such as the energy and angular spread of the produced electrons and the effective work function of the photocathode.
Section~\ref{simulations} discusses simulation results that were produced by Kassiopeia, a particle-tracking software that has been developed as a joint effort in the KATRIN collaboration over recent years~\cite{Kassiopeia2016}. These simulations allow us to gain a detailed understanding of the electron acceleration and transport processes inside the electron source.

\section{Setup and Design}
\label{design}

\subsection{Principle of the MAC-E filter}
\label{design:mace}

The principal design of the MAC-E filter is based on the combination of an electric retarding potential with a spatially inhomogeneous magnetic field~\cite{MACE1992}. In the following we describe this principle on the basis of the technical implementation at the KATRIN experiment.
Two solenoids located at the entrance and exit regions produce a strong magnetic field $B_\mathrm{max}$, which drops to a minimal value $B_\mathrm{min}$ at the central plane of the spectrometer. The value $B_\mathrm{min}$ can be adjusted by a system of air coils, which are placed around the spectrometer. The beam tube and the electrodes at the spectrometer entrance and exit are on ground potential, while the spectrometer vessel and the central electrodes are operated at high voltage. The absolute value of the \emph{retarding potential} increases towards the central spectrometer plane and reaches a maximum of $U_\mathrm{ana} \approx \SI{-18.6}{kV}$ at the position of the magnetic field minimum. This point lies on the so-called \emph{analyzing plane}.
The electromagnetic conditions in the analyzing plane define the transmission function for electrons that propagate through the spectrometer. Inside the MAC-E filter, electrons follow a cyclotron motion around the magnetic field lines. The kinetic energy $E$ can be split into a longitudinal component $\longit{E}$ into the direction of the field line and a transversal component $\transv{E}$, wich corresponds to the gyration around the field line.
Both components of the electron's kinetic energy can be described by the polar angle of the electron momentum relative to the magnetic field line, the \emph{pitch angle} $\theta = \angle{(\vec{p},\vec{B})}$:
\begin{equation}
    \label{eq:energy_trans_long}
    \transv{E}  = E \cdot \sin^2 \theta
    \,,
    \qquad
    \longit{E}  = E \cdot \cos^2 \theta
    \,.
\end{equation}

The adiabatic motion of the electrons is one of the key features of the MAC-E filter. When the relative change of the magnetic field over one cyclotron turn is small, the magnetic moment $\mu$ is conserved (here written non-relativistically):
\begin{equation}
    \label{eq:magnetic_moment}
    \mu = \frac{\transv{E}}{\magnit{B}} = \const
        \mtext{for}\quad
        \left| \dfrac{1}{B} \diff{B}{z} \right| \ll \dfrac{\omega_\mathrm{c}}{\longit{v}}
    \,,
\end{equation}
where $\omega_\mathrm{c}$ denotes the cyclotron frequency of the electron.
The reduction of the magnetic field towards the analyzing plane leads to a decrease in transversal energy $\transv{E}$. The longitudinal component $\longit{E}$ increases accordingly in this process because of energy conservation. This behavior results in a momentum collimation of the electron beam, and electrons that enter the spectrometer at a strong magnetic field $B_\mathrm{max}$ reach a minimal transversal energy $\transv{E}$ at $B_\mathrm{min}$ in the analyzing plane. Because the retarding potential $U_\mathrm{ana}$ only analyzes the longitudinal energy component, this principle of adiabatic collimation allows measuring the energy of electrons from an isotropic source with high precision.
The transmission condition for an electron with charge $q$ that enters the spectrometer with energy $E_0$ and pitch angle $\theta_0$ is
\begin{equation}
    \label{eq:transmission_cond}
    q U_\mathrm{ana} < \longit{E} = E_0 \cdot \left( 1 - \sin^2 \theta_0 \cdot \frac{B_\mathrm{min}}{B_\mathrm{max}} \right)
    \,.
\end{equation}
At nominal conditions, the KATRIN main spectrometer achieves a minimal magnetic field of $B_\mathrm{min} = \SI{0.3}{mT}$ in the analyzing plane and a maximal magnetic field of $B_\mathrm{max} = \SI{6}{T}$ at the pinch magnet, which is positioned at the exit of the main spectrometer.

The energy resolution (more precisely: filter width) $\Delta E$ of the MAC-E filter for an isotropic source is derived by postulating an electron that enters the spectrometer with maximum pitch angle, $\theta_0 = \ang{90}$ or $\transv[start]{E} = E_0$. The energy resolution corresponds to the remaining transversal kinetic energy in the analyzing plane after adiabatic collimation in the MAC-E filter:
\begin{equation}
    \label{eq:energy_resolution}
    \Delta E = \transv[max]{E} = E_0 \cdot \frac{B_\mathrm{min}}{B_\mathrm{max}} = \SI{0.93}{eV}
    \,.
\end{equation}
At KATRIN, electrons start in the source at $B_\mathrm{source} < B_\mathrm{max} = \SI{6}{T}$. This leads to \emph{magnetic reflection} of electrons with large pitch angles at the pinch magnet. The magnetic mirror effect occurs independently of the spectrometer transmission condition and reduces the acceptance angle of the MAC-E filter,
\begin{equation}
    \label{eq:magnetic_reflection}
    \theta_0 \le \theta_\mathrm{max} = \arcsin\sqrt{ \frac{B_\mathrm{source}}{B_\mathrm{max}} }
    \,,
\end{equation}
and electrons created with larger pitch angles do not contribute to the measurement.
At KATRIN, the acceptance angle is $\theta_\mathrm{max} = \ang{51}$ for electrons starting in the WGTS with $B_\mathrm{source} = \SI{3.6}{T}$. This excludes electrons with excessive path lengths as a result of their cyclotron motion, and thereby reduces systematic uncertainties caused by energy losses.

The KATRIN beam line transports a maximum magnetic flux of \SI{191}{T.cm^2} from the source to the detector. Electrons that are created at the source follow different magnetic field lines, depending on their initial radial and azimuthal position. The transmission function for electrons is affected by inhomogeneities in the analyzing plane of the electric potential ($\Delta U_\mathrm{ana} < \SI{1.2}{V}$) and the magnetic field ($\Delta B_\mathrm{min} < \SI{50}{\micro T}$). Because these variations are too large to be neglected, the detector features a pixelated wafer that can adequately resolve the position in the analyzing plane. This allows us to consider the electromagnetic inhomogeneities by determining transmission functions for individual detector pixels.
The exact value of the retarding potential $U_\mathrm{ana}$ and the magnetic field $B_\mathrm{min}$ can be accessed through measurements with an electron source that generates electrons at defined kinetic energy and pitch angle. A source that fulfills these requirements has been developed at WWU M{\"u}nster for the commissioning of the KATRIN main spectrometer.

\subsection{Principle of the electron source}
\label{design:principle}

\begin{figure}[tb]
        \includegraphics[width=\columnwidth]{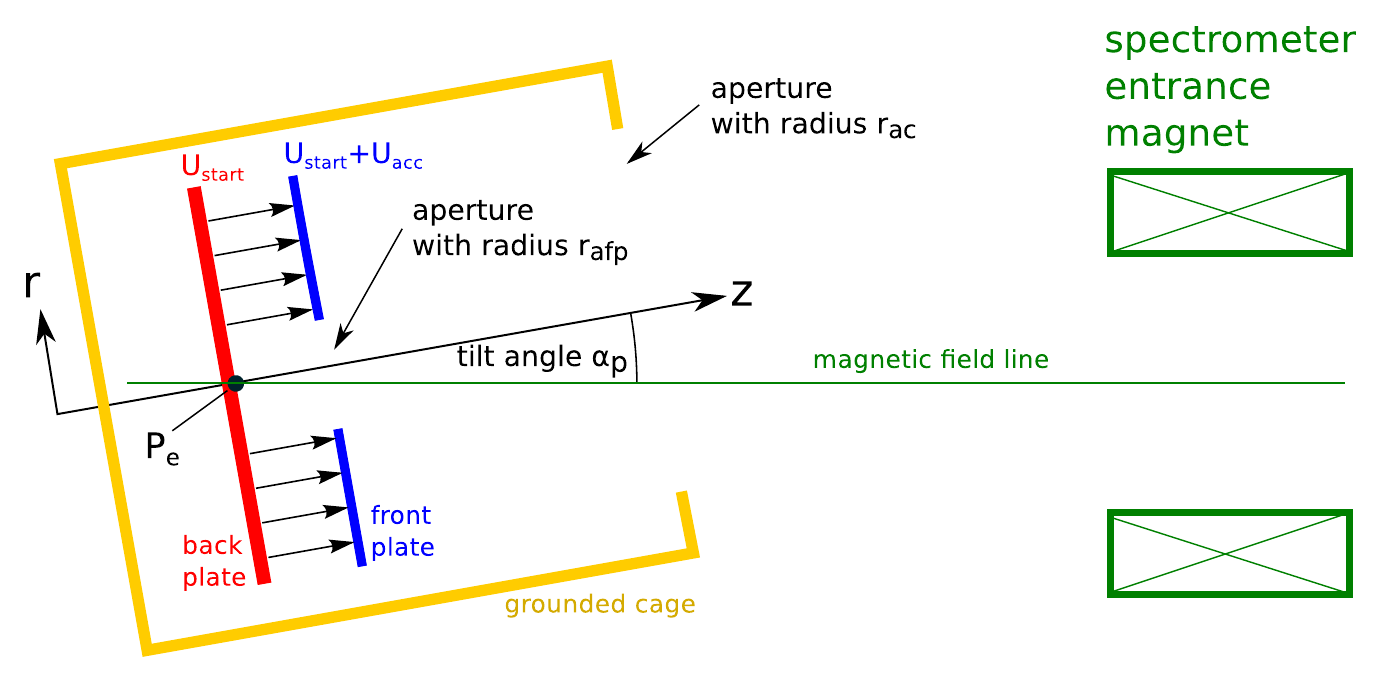}
        \caption{Schematic drawing of the electron source. The electrons are created by photo-emission from a thin photocathode layer and accelerated by a strong electric field inside a rotationally symmetric plate-capacitor setup. The grounded cage shields the electric acceleration field at the photocathode from outside influences.
        The complete setup can be tilted against the direction of the magnetic field to imprint a defined pitch angle on the generated electrons.}
    \label{fig:egun}
\end{figure}

It was demonstrated in \cite{Egun2011} that angular selectivity can be achieved by a combination of non-parallel electric and magnetic fields. An earlier design that used a gold-plated quartz tip, which was illuminated by UV light from optical fibers on the inside of the tip was able to produce electrons with non-zero pitch angles. This setup achieved an insufficiently large angular spread of the electrons. The source was therefore not usable as a calibration source for a MAC-E filter.
The design was further refined in \cite{Egun2014} and the setup now resembles a plate capacitor that introduces a homogeneous electric acceleration field. The setup can be tilted against the magnetic field lines to imprint a well-defined pitch angle on the generated electrons. This design uses a planar photocathode, which is back-illuminated by UV light from a single optical fiber.

This setup is shown in figure~\ref{fig:egun}. The emission spot $p_e$ is located on the \emph{back plate} (red), which is put on a negative potential $U_\mathrm{start}$ and thus defines the kinetic energy of the generated electrons, $E_\mathrm{kin} = q U_\mathrm{start}$. The \emph{surplus energy} of the electrons in the analyzing plane,
\begin{equation}
    \label{eq:surplus_energy}
    q \Delta U = q ( U_\mathrm{start} - U_\mathrm{spec} )
    \,,
\end{equation}
then amounts to the remaining kinetic energy that is available to overcome the retarding potential $U_\mathrm{ana}$ of the spectrometer.
The inhomogeneity $\Delta U_\mathrm{ana}$ of the retarding potential caused by the finite dimensions of the spectrometer is called \emph{potential depression}. It results in an effective retarding potential $U_\mathrm{ana} = U_\mathrm{spec} + \Delta U_\mathrm{ana}$ that is more positive than the \emph{spectrometer voltage} $U_\mathrm{spec}$. The value $U_\mathrm{ana}$ is affected by further inhomogeneities of the electromagnetic conditions in the spectrometer (\eg{} work function fluctuations), which can be resolved by transmission function measurements with our electron source.

The \emph{front plate} (blue) with an aperture for electrons is mounted parallel to the back plate and placed in front of the emission spot. A potential difference $U_\mathrm{acc} = U_\mathrm{front} - U_\mathrm{start} \le \SI{5}{kV}$ is applied between the plates to create an electric field perpendicular to the photocathode surface. The plates are mounted inside a grounded cage (yellow) to shield the electric field at the photocathode against outside influences. The whole setup can be mechanically tilted against the direction of the magnetic field. After passing the front plate, the electrons are accelerated adiabatically towards the ground potential at the spectrometer entrance where they achieve their maximum kinetic energy.

The electrons are emitted from a photocathode that consists of a thin gold (or silver) layer. The photocathode is back-illuminated via an optical fiber by UV light with a variable wavelength $\lambda$, which can be tuned to match the work function $\Phi$ of the photocathode material, $\lambda \lesssim hc / \Phi$. The energy distribution of the emitted electrons is defined by the photon energy $h \nu = hc / \lambda$ and the work function $\Phi$:
\begin{equation}
    \label{eq:photoeffect}
    0 < E_e  \le  h \nu - \Phi = hc / \lambda - \Phi
    \,.
\end{equation}
For metallic surfaces used in our electron source, the work function is $\Phi < \SI{5}{eV}$. Factors such as surface roughness or impurities caused by adsorbed gas molecules modify the Fermi level at the photocathode, which typically reduces the observed work function~\cite{Photoemission2012}. It is possible to perform an \emph{in situ} measurement of the work function using the well-known approach by Fowler \cite{Photoemission1931}. In section~\ref{measurements:workfunction} we present results from applying this technique.

In our electron source setup with a planar photocathode, the initial emission angle of the emitted electrons is expected to follow a $\cos\theta$-distribution~\cite{Photoemission2002}. A well-defined pitch angle $\theta$ is created by collimating the electron beam with the electric field $\vec{E}$ between the two plates. The electrostatic acceleration dominates the electron propagation because of their low kinetic energy after emission, according to the Lorentz equation
\begin{equation}
    \label{eq:lorentz}
    \vec{F} = q (\vec{E} + \vec{v} \times \vec{B})
    \,.
\end{equation}
The magnetic guiding field $\vec{B}$ takes over as the electrons gain more kinetic energy, and eventually the electrons enter an adiabatic cyclotron motion around the magnetic field line. The resulting pitch angle of the electrons in the spectrometer entrance magnet is minimal if the plate setup is aligned with the magnetic field, $\alpha_\mathrm{p} = \ang{0}$. By tilting the source against the magnetic field by the \emph{plate angle} $\alpha_\mathrm{p} > \ang{0}$, the non-adiabatic acceleration by the electric field works against the magnetic guiding field. This increases the transversal kinetic energy of the electrons, thereby creating an angular distribution of gaussian shape with a defined mean pitch angle $\theta > \ang{0}$. Because the plate setup is located inside a grounded cage, the electric acceleration field at the photocathode is constant for different plate angles.

The pitch angle of the emitted electrons transforms adiabatically during propagation towards the spectrometer entrance magnet, where the electrons enter a higher magnetic field. According to \eqref{eq:magnetic_moment} the transformation depends on the ratio of the magnetic fields at the emission spot, $B_\mathrm{start}$, and the magnetic field in the entrance magnet, $B_\mathrm{mag} \le B_\mathrm{max}$. The pitch angle increases because $B_\mathrm{start} \ll B_\mathrm{mag}$. The electron source we present here produces defined pitch angles that cover the full range of $\theta = \SIrange{0}{90}{\degree}$ in the entrance magnet with plate angles $\alpha_\mathrm{p} \le \ang{16}$ (section~\ref{measurements:transmission}).
The energy spread in the entrance magnet is defined by the initial energy distribution of the emitted electrons, because the acceleration by the electric field does not deform the energy distribution. The kinetic energy is merely shifted by $q U_\mathrm{start}$, while the spectral shape of the initial distribution is unaffected. A spectroscopic measurement of the electron energy, \eg{} with a MAC-E filter, therefore allows us to determine the initial energy distribution of the emitted electrons. The method is also suited to determine the photocathode work function, which is discussed in section~\ref{measurements:linewidth}.

\subsection{Technical design}
\label{design:egun}

The technical design of the electron source is based on the plate-capacitor setup depicted in figure~\ref{fig:egun}. We use two stainless steel disks with radius $r_\mathrm{p} = \SI{30}{mm}$ for the front and back plate, which are placed at a distance $d = \SI{10}{mm}$. Both plates were electro-polished before installation. The front plate has a thickness of $d_\mathrm{fp} = \SI{2}{mm}$ and features an aperture with a radius $r_\mathrm{afp} = \SI{3}{mm}$ for the emitted electrons. The back plate has a thickness of $d_\mathrm{bp} = \SI{3}{mm}$ and allows mounting a photocathode holder at its center. The holder has an aperture to glue-in an optical fiber with diameter \SI{200}{\micro m}. The holder with the optical fiber is manually polished to create a flat surface, and the photocathode material is deposited on the surface by {electron beam physical vapor deposition} (EBPVD). For the measurements presented here we used a gold photocathode with a layer thickness of \SI{20}{nm}; we also used silver with a thickness of \SI{40}{nm} in other measurements. The plates are isolated against each other and the grounded cage by {polyether ether ketone} (PEEK) insulators. The grounded cage has an inner radius of $r_\mathrm{c} = \SI{50}{mm}$ with an aperture $r_\mathrm{ac} = \SI{35}{mm}$ at the front.

The grounded cage is gimbal-mounted to allow tilting against two axes. The center of rotation is aligned with the emission spot on the back plate. This design ensures that the magnetic field line that the electron is following does not change when tilting the source cage. A precise readout of the plate angle is achieved by rotating piezo-electric motors (Attocube ANR240) that are installed at the pivot joints of the gimbal mount. These motors do not provide sufficient torque to tilt the electron source under vacuum conditions, but allow the relative tilt angle to be measured with a precision of \ang{0.05}.
To actuate the gimbal mount under vacuum conditions, our design uses two air-pressure linear motors (Bibus Tesla 1620) that are mounted outside the vacuum chamber. The linear motion of the motors is transferred onto the chamber by Bowden cables that are attached to each axis of the gimbal mount. By operating the motors, each axis can be tilted separately. The motors are controlled with a LabView software, which also takes care of the transformation between the two-axial and polar/azimuthal coordinate system for the plate angles.

When electrons are reflected by the electric retarding potential at the analyzing plane or by the magnetic field at the spectrometer entrance, they may become stored between the spectrometer and the electron source in a setup similar to a Penning trap. This can lead to a discharge, which has disastrous consequences for the photocathode. To avoid such storing conditions and the subsequent discharge, a dipole electrode is placed in the beamline, between the source cage and the entrance magnet. The electric field $\vec{E}$ induced by the electrode results in a drift of the stored electrons,
\begin{equation}
    \label{eq:ExB}
    \vec{v}_\mathrm{drift} = \dfrac{\vec{E} \times \vec{B}}{B^2}
    \,,
\end{equation}
with $\vec{B}$ the magnetic field at the dipole electrode. In our setup, the half-shell dipole electrode spans \ang{170} at a radius of $r_\mathrm{dip} = \SI{30}{mm}$ and is operated at a voltage $U_\mathrm{dip} \le \SI{4}{kV}$.
Measurements confirmed that this electrode removes trapped electrons efficiently and prevents Penning discharges; this is discussed in section~\ref{measurements:rates}.

The optical system to provide the UV light for the photocathode allows choosing between two light sources.
A frequency-quadrupled $\text{Nd:YVO}_4$ laser (InnoLas mosquito-266-0.1-V) provides UV light at a wavelength of \SI{266}{nm} (\SI{1}{nm} FWHM) at high intensity ($< \SI{10}{mW}$ output power). The intensity can be adjusted by an internal attenuator ($\lambda/2$-plate with polarizing filter) and by a neutral density (ND) filter, which is placed in the laser beam.
Behind the ND filter, a fraction of approximately 0.5\% of the UV light is coupled out by a beam splitter to measure the UV light intensity with a photodiode. The laser light is focused by an aspheric lens into a $\varnothing \SI{200}{\micro m}$ optical fiber and guided into the source chamber. The laser is operated in pulsed mode with frequencies of \SIrange{40}{100}{kHz} at a pump diode current of \SIrange{6}{8}{A}. The current and frequency setting determines the output power, which can be tuned to produce a desired electron rate of several \si{kcps} (\si{cps}: counts per second) at the detector. The pulse width of $< \SI{20}{ns}$ allows for time-of-flight measurements with a precisely known starting time of the electrons.

Alternatively, an array of LEDs can be used as light source to provide UV light with $\lambda = \SIrange{260}{320}{nm}$. Six ball-lens UV LEDs (Roithner UVTOP260--310) with peak wavelengths of \SIlist{265;275}{nm} \mbox{etc.} on are mounted on a revolver that is moved by a stepper motor. This allows us to automatically place the desired LED on the optical axis without manual adjustments. To achieve a sharp line width, a UV monochromator with \SI{4}{nm} FWHM is used. The monochromator is operated by another stepper motor. The LED revolver in combination with the monochromator allows selecting arbitrary wavelengths in the available range.
Like in the laser setup, a beam splitter with photodiode is used to monitor the light intensity. The divergent light beam of the LEDs is focused by an optical telescope consisting of two convex lenses, and guided into the electron source through an optical fiber. The current to operate the LEDs is provided by a function generator in pulse mode, using the internal \SI{50}{\ohm} resistor with an output voltage of \SI{8.5}{V}. With this setting, the LEDs are driven by a peak current of \SI{200}{mA}, which corresponds to a mean current of \SI{20}{mA} at 10\% duty cycle. Under nominal conditions, a pulse frequency of \SI{100}{kHz} is used with a pulse length of \SI{1}{\micro s}. Time-of-flight measurements are thus also possible with LEDs as a light source. Depending on the LED and the monochromator setting, electron rates in the \si{kcps} range can be achieved.

The optical system (laser device, the stepper motors of the LED system and the two photodiodes), the actuation of the plate angle and the power supply for the dipole electrode are controlled and monitored by a LabView software that has been developed for use with the electron source.
The photodiode read-out allows us to monitor the stability of the UV light source, where intensity changes (\eg{} because of warm-up effects) could result in fluctuations of the observed electron rate.

\subsection{Analytical transmission function}
\label{design:transmission}

The observed transmission functions from measurements with the electron source can be modeled by an analytical description of the MAC-E filter~\cite{groh:phd}. The conditions for transmission \eqref{eq:transmission_cond} and magnetic reflection \eqref{eq:magnetic_reflection} are applied to the theoretical energy distribution $\eta(E)$ and angular distribution $\zeta(\theta)$. The analytical transmission function $T(E)$ is given by the integrated energy distribution, which is modified by the range of pitch angles that are transmitted through the spectrometer:
\begin{equation}
    \label{eq:transmission_model}
    T(E, U_\mathrm{ana})
            = \dot{N}_\mathrm{0} \cdot \int_{E}^{\infty} \eta(\eps) \; \int_{0}^{\theta_\mathrm{max}(\eps, U_\mathrm{ana})} \zeta(\theta) \; \dx{\theta} \dx{\eps} \; + \dot{N}_\mathrm{b}
    \,,
\end{equation}
where $\dot{N}_\mathrm{0}$ is the amplitude of the electron signal and $\dot{N}_\mathrm{b}$ the observed background. The term $\theta_\mathrm{max}$ describes the largest pitch angle that can be transmitted according to \eqref{eq:transmission_cond}:
\begin{equation}
    \theta_\mathrm{max}(E, U_\mathrm{ana}) =
        \begin{cases}
            \arcsin\left( \sqrt{ \frac{E}{U_\mathrm{ana}} \cdot \frac{2}{\gamma + 1} \cdot \frac{B_\mathrm{start}}{B_\mathrm{min}} } \right)
        \,,
        \\
            0 \qquad\mtext{for} \sqrt{\ldots} < 0
        \,.
        \end{cases}
\end{equation}
This analytical method includes all relevant effects into the model (\eg{} the transformation of the pitch angle resulting from adiabatic collimation), and allows us to determine the underlying distributions independently~\cite{erhard:phd,behrens:phd}.

The asymmetric energy distribution is described by a generalized normal distribution~\cite{GeneralizedGaussian1997},
\begin{equation}
    \label{eq:transmission_model_energy}
    \eta(E)     =  \frac{1}{\sqrt{2 \pi}} \cdot
                    \begin{cases}
                        \frac{1}{\alpha_E} \cdot
                            \exp\left( -\onehalf \frac{(E - \hat{E})^2}{\alpha_E^2} \right)
                        &(\kappa = 0)
                    \,,
                    \\
                        \frac{1}{\alpha_E - \kappa (E - \hat{E})} \cdot
                    \\
                        \quad   \cdot \exp\left( -\frac{1}{2 \kappa^2} \ln\left[ 1 - \kappa \frac{E - \hat{E}}{\alpha_E} \right]^2\right)
                        &(\kappa \ne 0)
                    \,,
                    \end{cases}
\end{equation}
with the mean energy $\hat{E}$ and the energy width $\alpha_E$. For the transmission model, the energy distribution is evaluated in the range $E = [0; \infty)$.
The asymmetry is described by the skewness parameter $\kappa$; at $\kappa = 0$ the distribution is equivalent to a symmetric normal distribution. For $\kappa > 0$, the function is limited to $E = [0; \hat{E} + \frac{\alpha_E}{\kappa})$.
The width $\alpha_E$ can be converted into an energy spread $\sigma_E$, which can be compared independently of the skewness:
\begin{equation}
    \sigma_E    = \frac{\alpha_E}{\kappa} \cdot \sqrt{ e^{\kappa^2} ( e^{\kappa^2} - 1 ) }
    \,.
\end{equation}

The angular distribution is modeled by the sum of two normal distributions that are placed around $\theta = \ang{0}$,
\begin{equation}
    \label{eq:transmission_model_angle}
    \zeta(\theta)   =  \frac{1}{\sqrt{2 \pi} \sigma} \cdot \left[
                                \exp\left( -\frac{(\theta - \hat{\theta})^2}{2 \sigma_\theta^2} \right) +
                                \exp\left( -\frac{(\theta + \hat{\theta})^2}{2 \sigma_\theta^2} \right)
                            \right]
    \!\!,
\end{equation}
with the mean angle $\hat{\theta}$ and the angular spread $\sigma_\theta$. For the transmission model, the angular distribution is evaluated in the range $\theta = [\ang{0}; \ang{90}]$.
The summing takes into account that the distribution is deformed for $\theta \rightarrow \ang{0}$ because the pitch angle is only defined for positive values.

The measured transmission functions presented in this paper have been fitted by a Markov-Chain Monte Carlo (MCMC) method of minimizing the $\chi^2$ value, using a code that was implemented in Python. It utilizes \emph{emcee}~\cite{emcee2013} for the MCMC fit process~\cite{behrens:phd}.

\section{Measurements}
\label{measurements}

\begin{figure*}[tb]
        \centering
        \sidecaption
        \includegraphics[width=.6\textwidth]{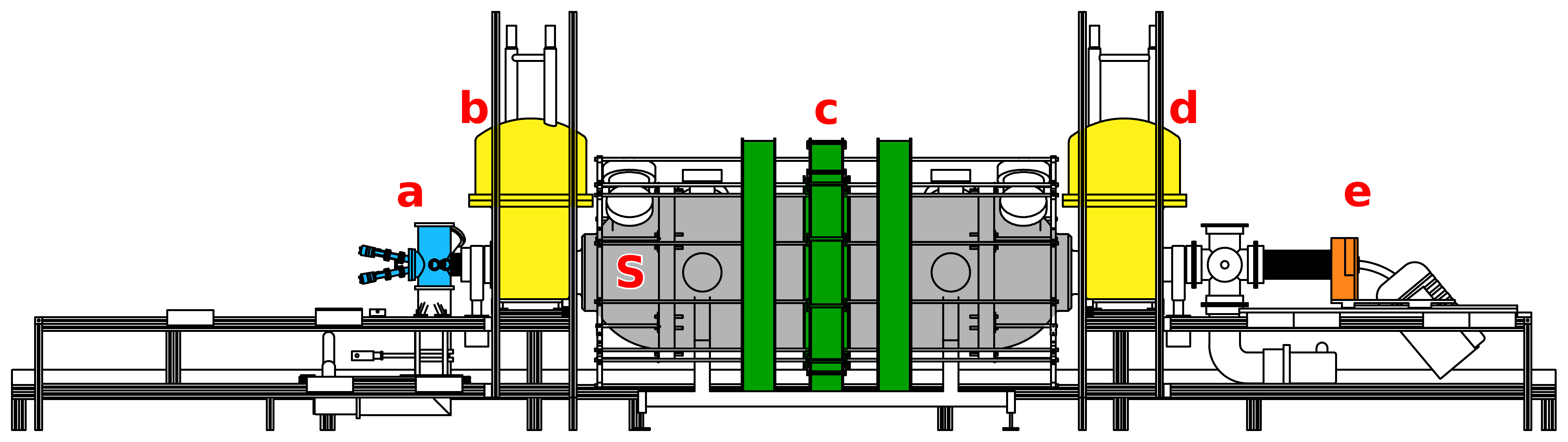}
        \caption{Test setup at the KATRIN monitor spectrometer. The electron source (a) is mounted in a vacuum chamber, which is connected to the spectrometer beamline. The spectrometer (S) is operated at high voltage up to \SI{-18.6}{kV} and follows a symmetric design with two solenoids (b,d) and four air coils (c) to adjust the magnetic field at the center. The electrons are detected by a \LN{}-cooled PIN diode (e).
        Details on the monitor spectrometer can be found in \cite{MonSpec2014}.
        }
    \label{fig:monspec-egun}
\end{figure*}

The photoelectron source presented in this work has been commissioned successfully at the KATRIN monitor spectrometer. The corresponding measurements were carried out in the summer of 2014 and allowed us to verify the two key features of the electron source -- angular selectivity and a small energy spread -- and to study other important characteristics of the device. The monitor spectrometer was chosen because it could be operated independently of the main spectrometer during hardware preparations for its second commissioning phase. The electron source was subsequently mounted at the main spectrometer for the commissioning measurements of the spectrometer and detector section~\cite{erhard:phd,behrens:phd,kraus:phd,barrett:phd,wierman:phd}.

\subsection{Experimental setup}
\label{measurements:setup}

\begin{figure}[tb]
        \includegraphics[width=.5\columnwidth]{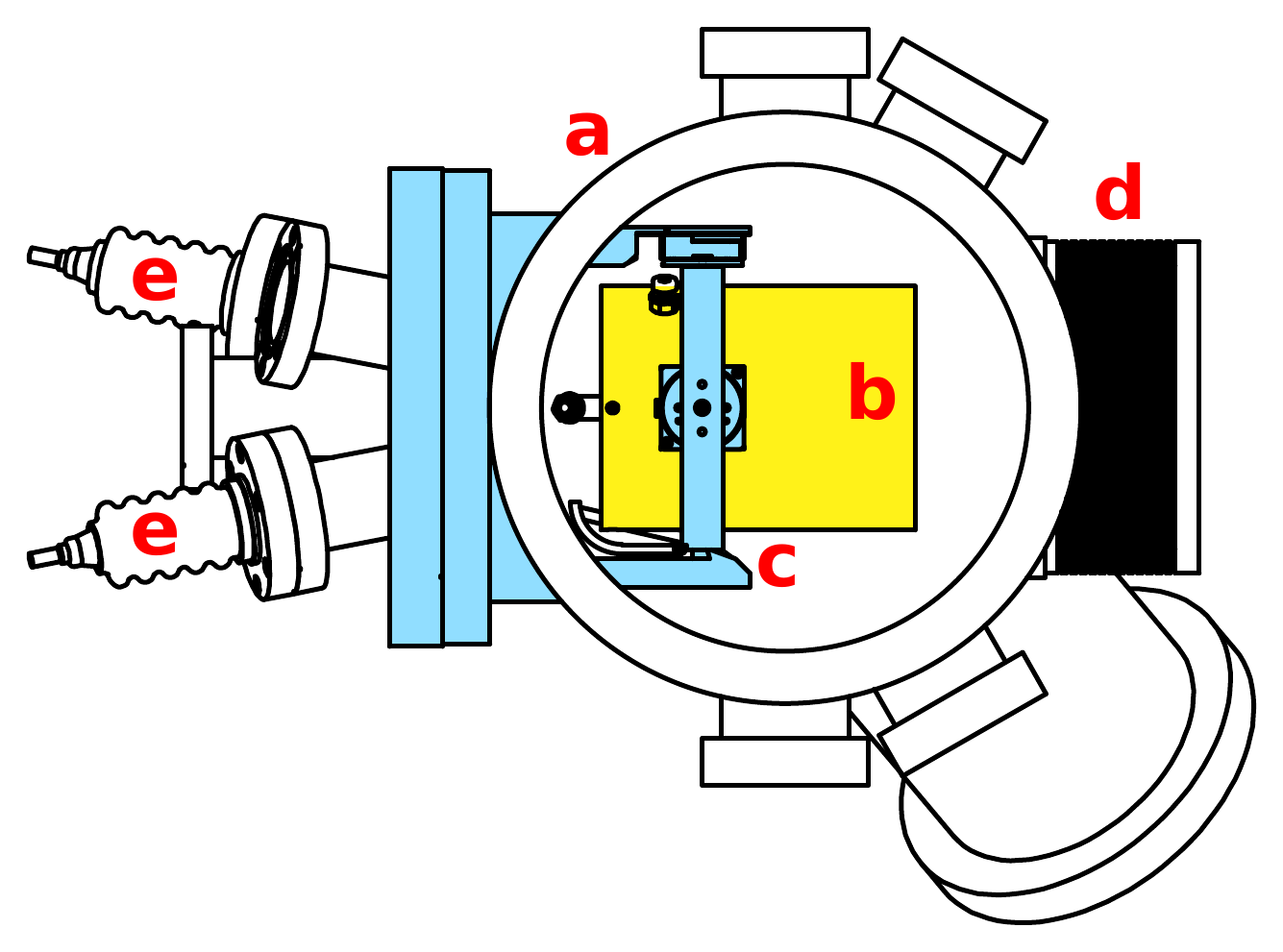}%
        \includegraphics[width=.5\columnwidth]{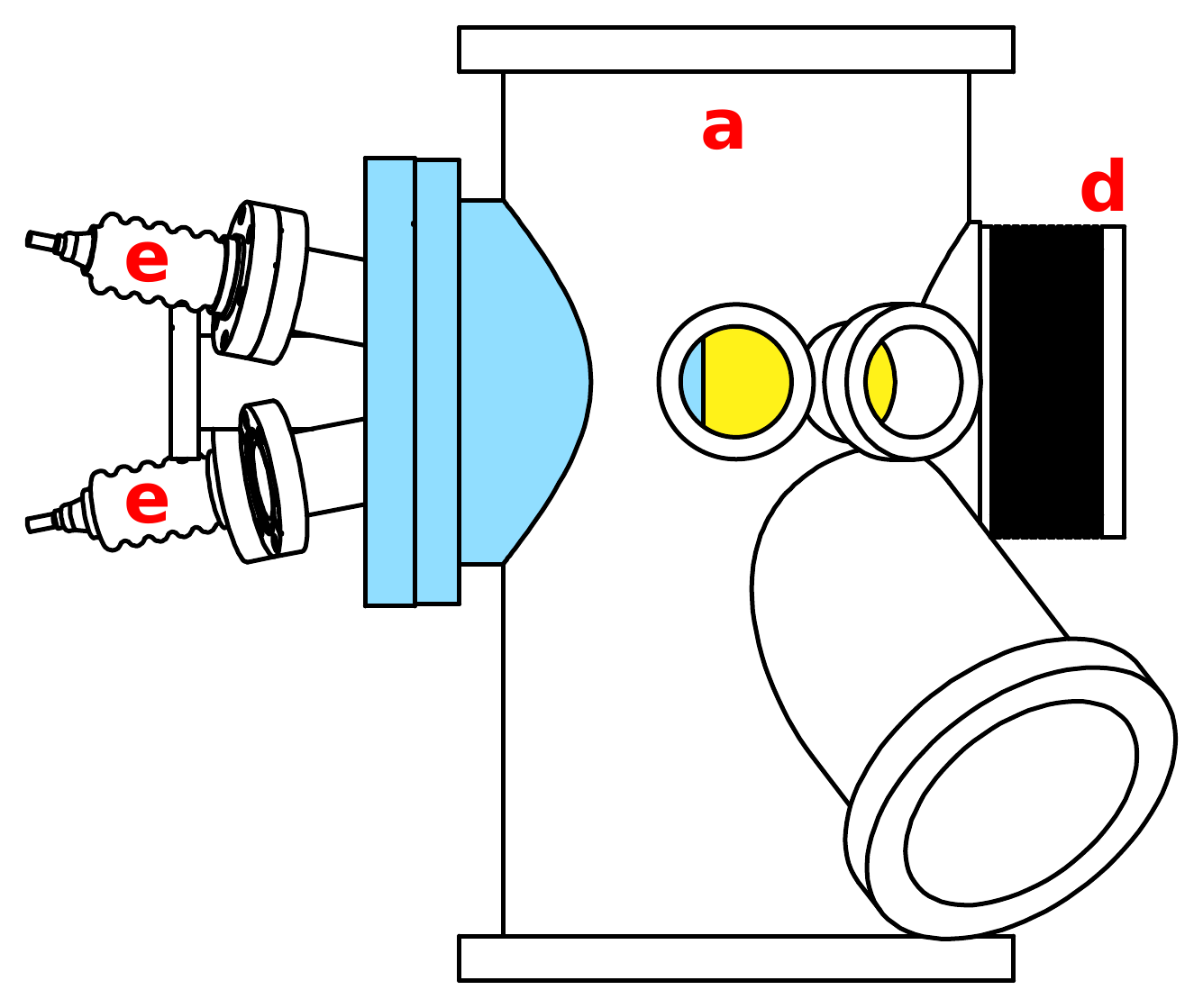}
        \caption{Detail of the test setup at the KATRIN monitor spectrometer. The electron source is mounted inside the vacuum chamber (a). The grounded cage (b) contains the plate-capacitor setup and can be tilted on a gimbal mount (c). A dipole electrode is located inside the bellow (d) that connects the vacuum chamber with the spectrometer beamline. The back and front plate are connected to a high voltage source via vacuum feed-throughs (e).}
    \label{fig:monspec-egun-elli}
\end{figure}

In contrast to the main spectrometer, the monitor spectrometer features a symmetric magnetic field setup with $B_\mathrm{max} = \SI{6}{T}$ at the spectrometer entrance and exit. In our measurements, the spectrometer was operated at voltages $U_\mathrm{ana} \approx \SI{-18.6}{kV}$ with a minimal magnetic field $B_\mathrm{min} = \SI{0.38}{mT}$ in the analyzing plane.
The electron source was mounted in a vacuum chamber in front of the spectrometer beamline. The full setup is shown in figs.~\ref{fig:monspec-egun} and \ref{fig:monspec-egun-elli}. The source was installed at a fixed position where the electron beam is always central to the spectrometer axis, as it was not necessary to perform measurements on different magnetic field lines.
The emission spot at the back plate is located at $z_\mathrm{es} = \SI{2.635}{m}$ in the spectrometer coordinate system, where the analyzing plane is located at $z_\mathrm{ana} = \SI{0}{m}$. The two solenoid magnets are placed symmetrically around the analyzing plane at an axial distance of $z_\mathrm{mag} = \pm \SI{2.01}{m}$; the reference point $z_0 = \SI{0}{m}$ refers to the analyzing plane at the spectrometer center.
The magnetic field at the emission spot was measured with a Hall probe, yielding a value $B_\mathrm{start} = \SI{21}{mT}$. To achieve electromagnetic conditions that are comparable to the main spectrometer setup, it is important to adjust $B_\mathrm{start}$ to achieve the magnetic field the electron source will be exposed to. Using an additional air-cooled coil that is placed close to the source at $z_\mathrm{coil} = \SI{-2.640}{m}$ and operated at a coil current $I_\mathrm{coil} = \SI{35}{A}$, the field at the emission spot increases to $B^+_\mathrm{start} = \SI{27}{mT}$. This is comparable to the main spectrometer setup where \SI{29}{mT} are reached.

\begin{figure}[tb]
        \includegraphics[width=\columnwidth]{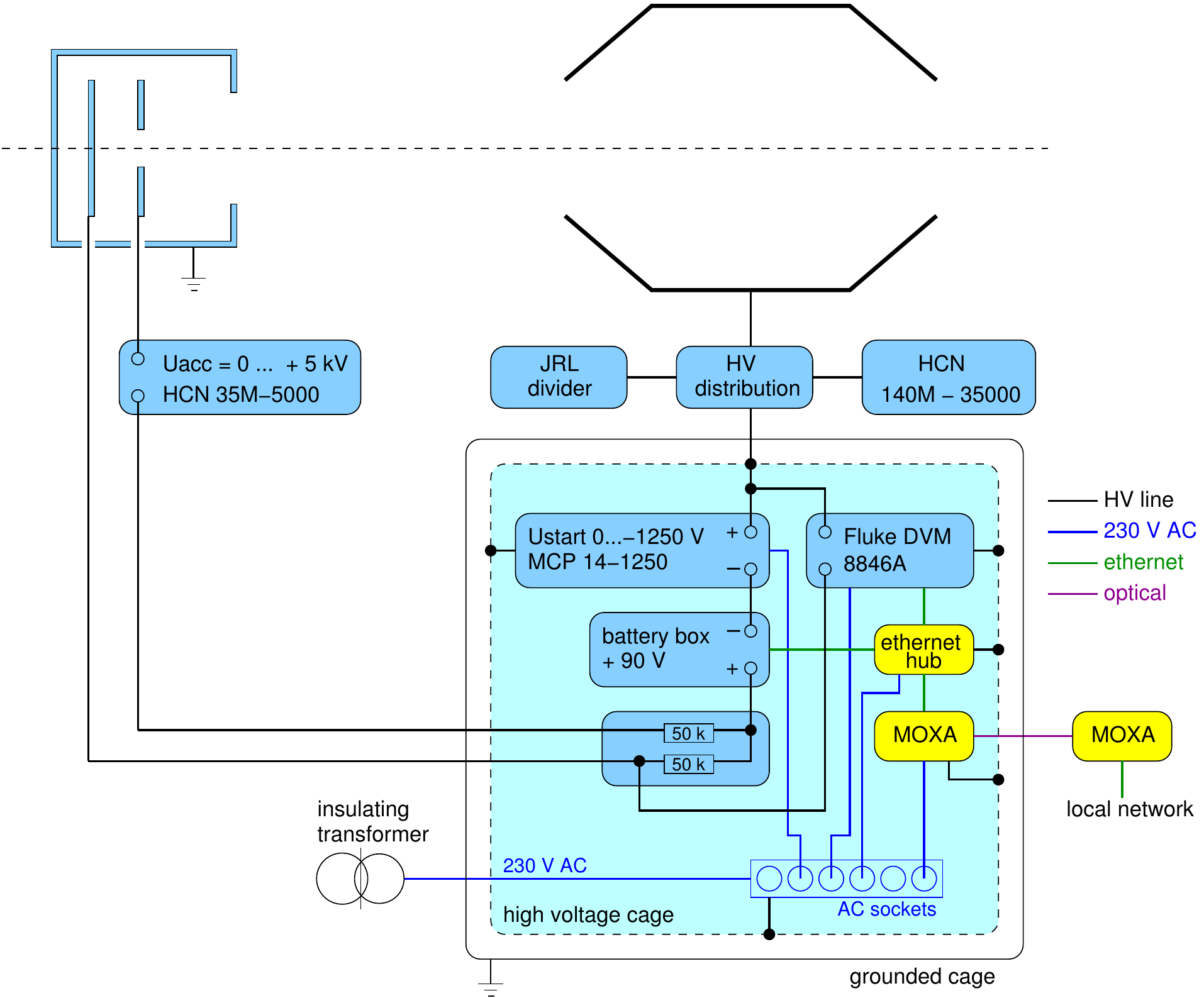}%
        \caption{High voltage scheme of the test setup. The base voltage provided to the spectrometer defines the retarding potential (HCN 140M-35000). The high voltage system of the electron source is placed inside a HV cabinet. The source voltage can be varied by means of a variable power supply (FuG MCP 14-1250) and a constant voltage source (battery). The resulting potential difference between electron source and spectrometer is monitored by a DVM (Fluke 8846A). The acceleration voltage is provided by an additional power supply (FuG HCN 35M-5000), which operates on top of the source voltage. The devices are controlled by a network interface; the connection into the HV cabinet is provided by a MOXA system with optical connections.}
    \label{fig:monspec-hv}
\end{figure}

Figure~\ref{fig:monspec-hv} shows the high voltage scheme of the monitor spectrometer setup. The electron source is connected with a small difference voltage to the high voltage of the spectrometer in order to cancel out voltage fluctuations that would occur if two independent power supplies were used. The back plate voltage, $U_\mathrm{start}$, can be varied against the spectrometer voltage $U_\mathrm{spec}$ by combining a power supply that operates at \SIrange{0}{-1.25}{kV} with a battery that delivers a voltage of about \SI{+90}{V}. By putting the two voltage sources in series, it is possible to vary the starting voltage to achieve a \emph{surplus energy} $q \Delta U = q (U_\mathrm{start} - U_\mathrm{spec}) = \SIrange{-90}{1160}{eV}$ without requiring a polarity-switching power supply. The voltage difference between electron source and spectrometer, $U_\mathrm{start} - U_\mathrm{spec}$, is measured by a difference voltmeter (DVM) to monitor the electron surplus energy. Transmission functions can be measured by varying the starting voltage within a few \si{V} around zero while observing the electron rate at the detector.
The high voltage system is mainly located inside a Faraday cage, which is operated on the spectrometer high voltage. This cage is put inside another grounded HV cabinet to allow safe operation.
The acceleration voltage for the front plate, $U_\mathrm{acc}$, is provided by an additional power supply that generates up to \SI{5}{kV} \wrt{} the back plate voltage. The acceleration voltage is thus kept constant while varying $U_\mathrm{start}$. This power supply is isolated for voltages up to \SI{35}{kV} and can be placed outside the HV cabinet.

\subsection{Electron rate}
\label{measurements:rates}

As a first test that the electron source is operating as expected, the achieved electron rate at the detector was determined. The electron rate depends on the UV light setting.
For the test measurements at the monitor spectrometer, we used a nominal laser setting of $f_\mathrm{las} = \SI{100}{kHz}$ and $I_\mathrm{las} = \SI{6}{A}$ with a 1\% ND filter in the optical beamline. This yields an electron rate of about $\dot{N} = \SI{1500}{cps}$ at the detector in full transmission. The statistical uncertainty of a \SI{10}{s} measurement is $< \SI{1}{\%}$ in this case, which is sufficient for our investigations and allows us to measure a typical transmission function in less than 10~minutes.
For some measurements, the light source was switched to LEDs to allow wavelength-dependent measurements. Because the electron emission is influenced by the photon energy (UV wavelength) and the work function of the photocathode material, the electron rate that is achieved with LEDs varies in a typical range of \SIrange{200}{1000}{cps}. The rate can be tuned over a wide range by changing the duty cycle of the function generator (pulser) that drives the LEDs.
The LEDs were typically operated at a pulse frequency $f_\mathrm{LED} = \SI{100}{kHz}$, a pulse width $\tau_\mathrm{LED} = \SI{1000}{ns}$ (10\% duty cycle), and a forward voltage $U_\mathrm{LED} = \SI{8.5}{V}$.

\begin{figure}[tb]
        \includegraphics[width=\columnwidth]{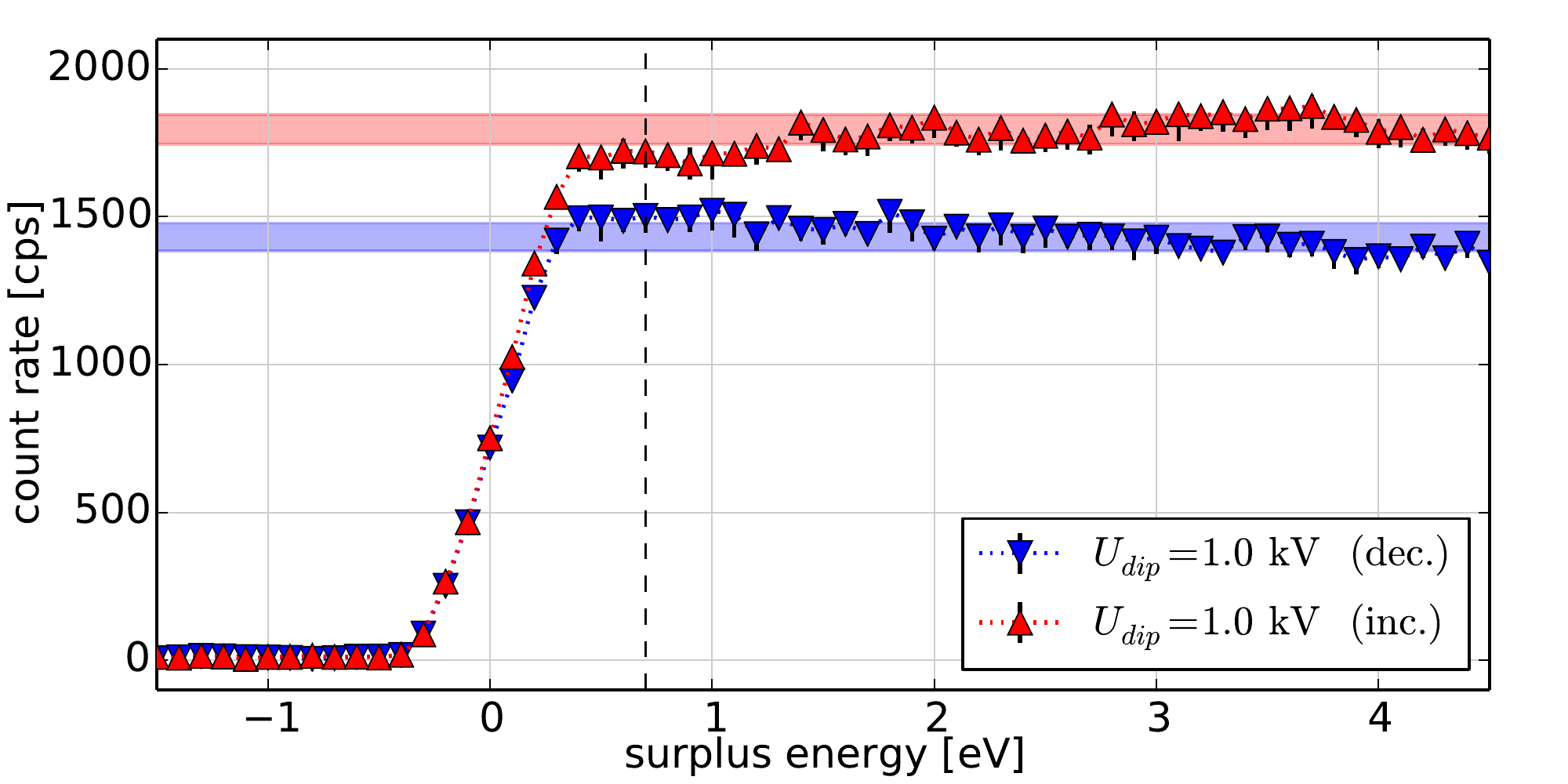}\\%
        \includegraphics[width=\columnwidth]{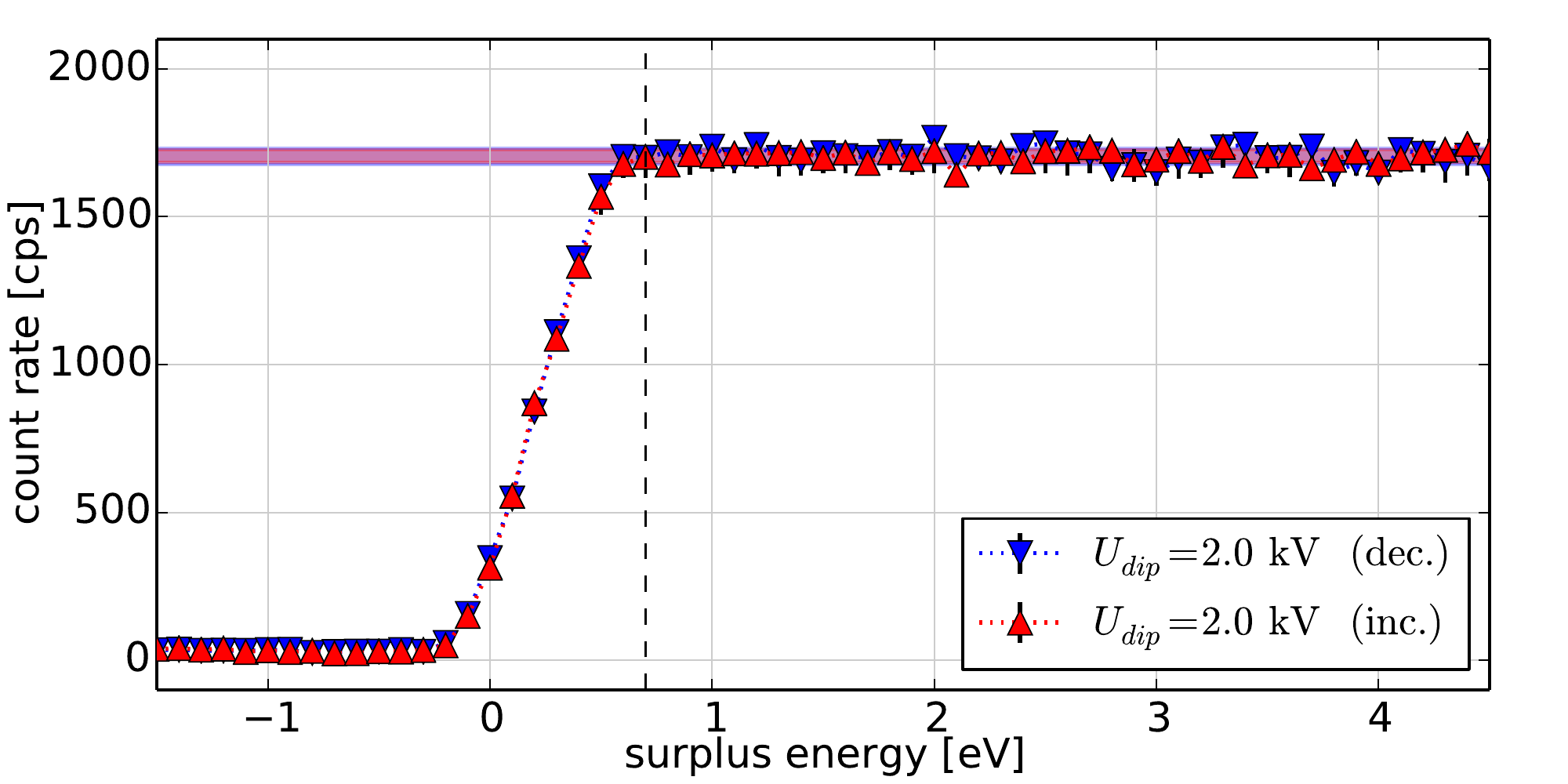}%
        \caption{Measured transmission functions at two different dipole voltages. If the dipole voltage is not sufficient to remove trapped electrons, a hysteresis effect is observed between measurements with decreasing ($\bigtriangledown$) or increasing ($\bigtriangleup$) surplus energy. The horizontal bars indicate the nominal rate in each measurement. The nominal rate is computed as the average of the respective data points on the right side of the dashed line. The bar width indicates the standard deviation of the averaged data points.
        While at $U_\mathrm{dip} = \SI{1}{kV}$ the hysteresis is clearly visible, it disappears completely at $U_\mathrm{dip} = \SI{2}{kV}$.}
    \label{fig:transmission-rate}
\end{figure}

The dipole electrode in front of the electron source is intended to remove stored electrons from the beamline between source and spectrometer. The removal efficiency depends on the strength of the induced \ExB{} drift \eqref{eq:ExB}, and thus increases with a larger dipole voltage\footnote{The alternative solution of reducing the global magnetic field would change the magnetic fields at the MAC-E filter and is thus disfavored.}.
In our setup, the magnetic field at the electrode is $B_\mathrm{dip} = \SI{78}{mT}$ with $E_\mathrm{dip} \approx \SI{40}{kV/m}$ according to simulations. The removal efficiency of the dipole electrode was investigated by measuring transmission functions in direction of increasing and decreasing electron surplus energy.

Figure~\ref{fig:transmission-rate} shows that the observed transmission function is affected by a hysteresis effect that depends on the dipole voltage, which allows investigating the removal efficiency of the dipole electrode. The observed transmission functions show a similar behavior, except for the nominal electron rate that is reached at full transmission. The small rate drift that can be observed in the upper panel can be explained by fluctuations in UV light intensity.
The hysteresis effect can be explained by the continuous filling of the trap from the beginning of the measurement when measuring in direction of increasing surplus energy, because the surplus energy at the beginning is too small for electrons to be transmitted. Electrons with a given energy stay trapped until they lost kinetic energy (\eg{} through synchrotron radiation) or are removed by the dipole field. Scattering processes with electrons of higher kinetic energy that are generated at a later time during the measurement cause some of the trapped electrons to gain kinetic energy, thereby increasing transmission probability towards the detector. The effect does not occur when the measurement is performed in inverse direction, where the higher-energetic electrons are transmitted at the beginning of the measurement~\cite{wierman:phd}.
This leads to a hysteresis effect in the electron rate between the two scanning directions, which becomes smaller when the dipole voltage is increased and more electrons are removed from the trap. The observed rate difference is therefore a direct measure for the dipole efficiency. Our measurement indicates that a dipole voltage of $U_\mathrm{dip} = \SI{2}{kV}$ is sufficient to avoid the hysteresis effect. With lower dipole voltages, the observed rate difference between the two scanning directions increases, indicating an insufficient removal of stored electrons.

\subsection{Energy spread}
\label{measurements:linewidth}

The energy resolution of a MAC-E filter \eqref{eq:energy_resolution} depends on the retarding potential $U_\mathrm{ana}$. At low voltages $|U_\mathrm{ana}| \ll \SI{18.6}{kV}$ and low electron energies $E \approx q U_{ana}$, the energy resolution improves because of the smaller amount of transversal energy left in the analyzing plane. A low voltage measurement with $U_\mathrm{ana} \approx \SI{-200}{V}$ allows us to directly determine the energy distribution of the produced electrons. Unfortunately, at the monitor spectrometer it is not possible to detect electrons with $E \ll \SI{10}{keV}$ due to the energy threshold of the detector. Fortunately, the energy distribution can also be determined from a measurement performed at nominal high voltage.

\begin{figure}[tb]
    \includegraphics[width=\columnwidth]{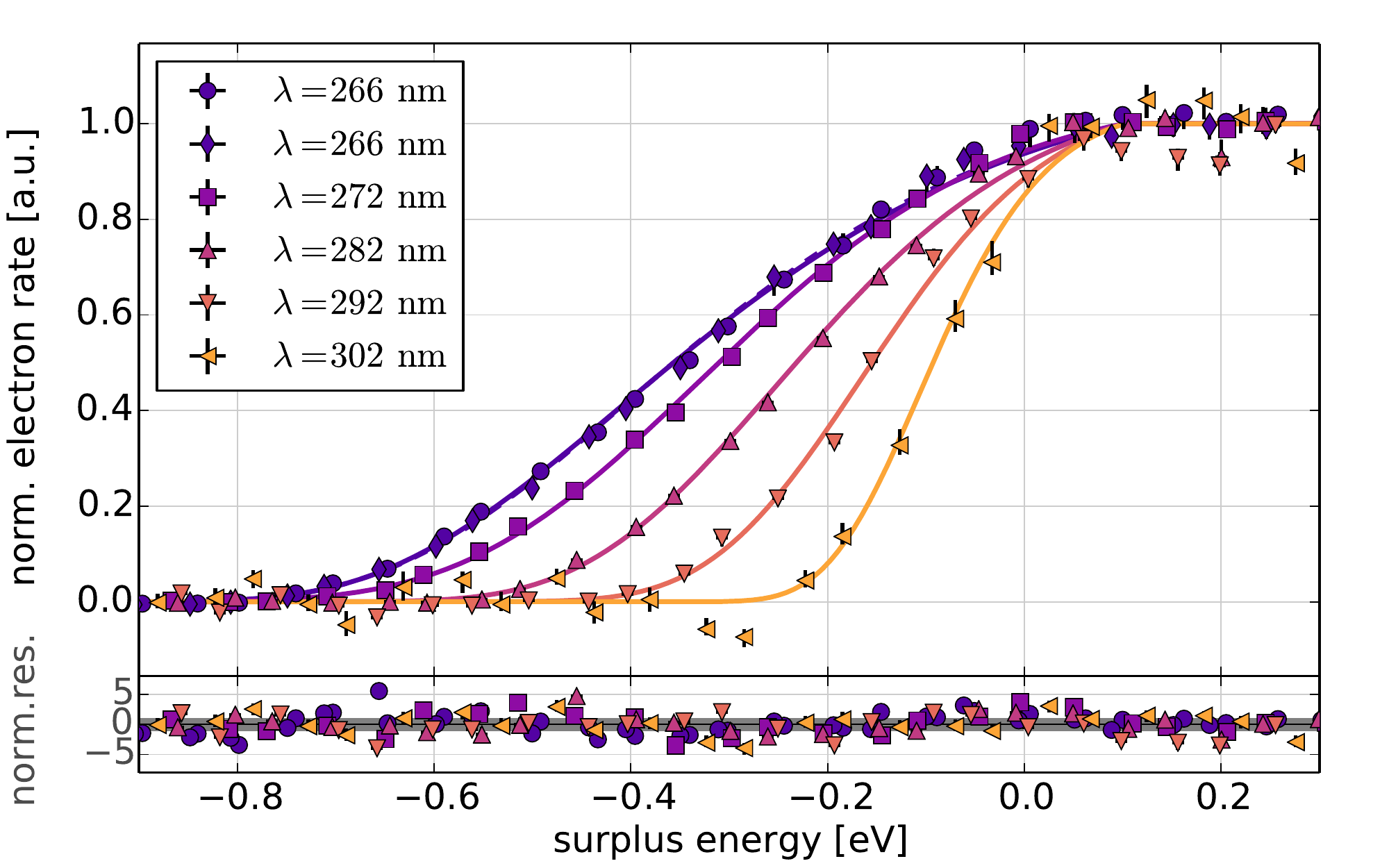}\\%
    \includegraphics[width=\columnwidth]{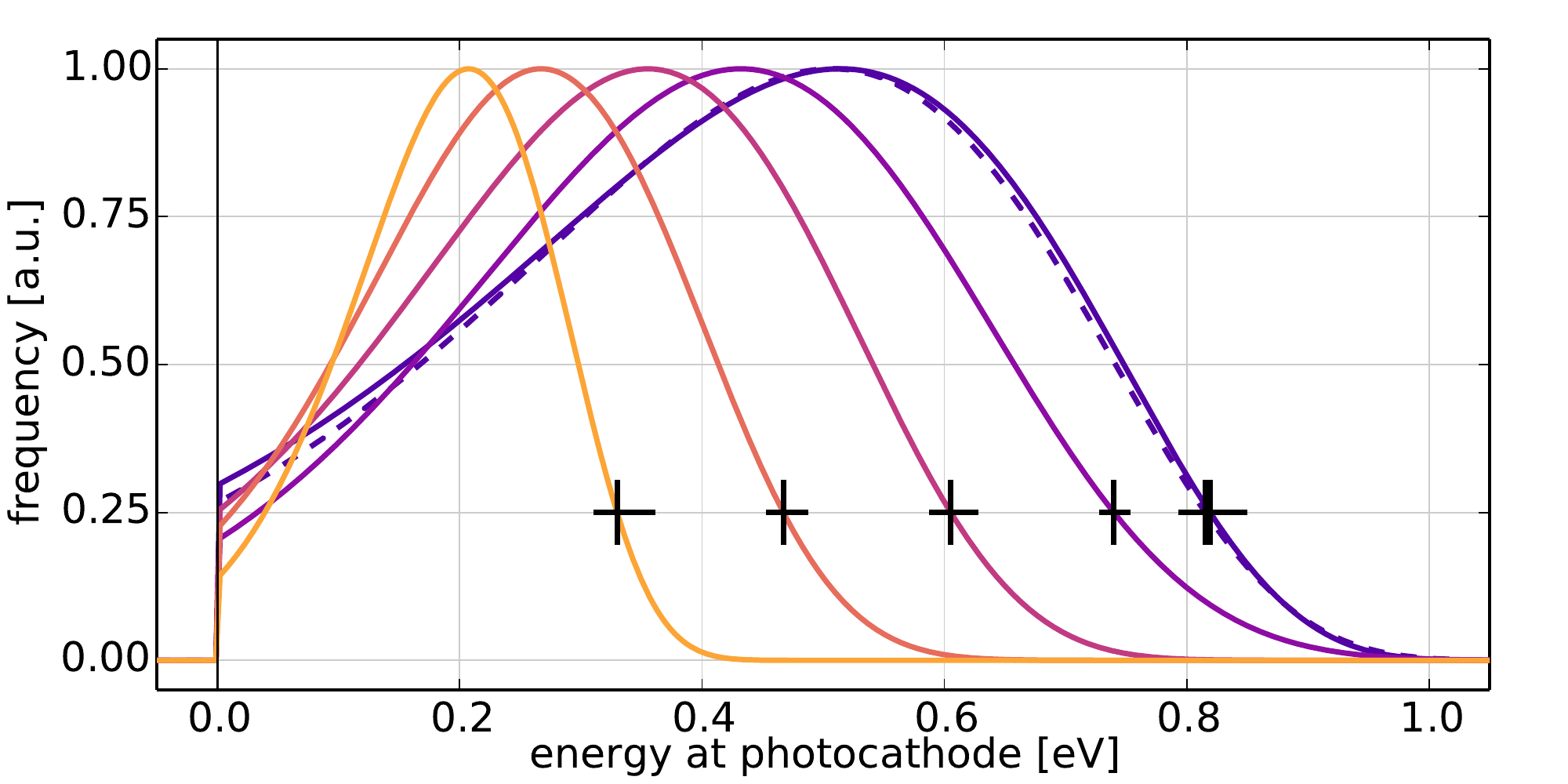}%
    \caption{Measured transmission functions and energy distributions with different UV light sources (laser: \SI{266}{nm}, LEDs \SIlist{275;285;295;305}{nm}) at $U_\mathrm{ana} = \SI{-18.6}{kV}$ and $B_\mathrm{start}^+ = \SI{27}{mT}$. The measurements (top panel) were performed at $\alpha_\mathrm{p} = \alpha_0$ (zero angle). The given residuals are normalized to the respective uncertainties of the data points; the gray band here indicates the $1\sigma$-limit.
    In this setting, the transmission functions are dominated by the energy distribution of the emitted electrons (lower panel). The dashed lines indicate a second measurement at same settings for $\lambda = \SI{266}{nm}$.
    The cross-shaped markers in the energy spectrum indicate the determined energy limit $E_\mathrm{max}$ and its uncertainty.
    Note that $E = \SI{0}{eV}$ in the energy distribution corresponds to the upper edge of the transmission function, where the observed transmission functions reach their nominal amplitude.}
    \label{fig:linewidth}
\end{figure}

If the source is operated at the so-called \emph{zero angle} setting, it produces the smallest possible pitch angle in the spectrometer entrance magnet. The zero angle position has to be found manually by varying the plate angle $\alpha_\mathrm{p}$ around \ang{0} independently for the vertical and horizontal axis. At an electron surplus energy $E = q \Delta U \approx \SI{0}{eV}$ \eqref{eq:surplus_energy}, the transmission probability is entirely dominated by the pitch angle of the emitted electrons. The observed electron rate is thus sensitive to small changes of the produced pitch angle, and the rate dependency \wrt{} the plate angle shows a maximum at zero angle $\alpha_\mathrm{p} = \alpha_0$. At the monitor spectrometer, the zero angle offset was found to be $\alpha_\mathrm{hor} = \SI{0.04(1)}{\degree}$ and $\alpha_\mathrm{ver} = \SI{1.13(1)}{\degree}$ at $U_\mathrm{dip} = \SI{2}{kV}$.
This offset is caused by mechanical imperfections, which result in a minor misalignment that can be easily corrected by such a measurement. The impact of the angular spread on the observed transmission function is marginal when the zero angle is applied. In this case the actual mean and width of the angular distribution are not relevant to the analytical transmission model as long as $\theta < \ang{5}$, and the energy spread dominates the shape of the resulting transmission function. It is thus possible to fit an (integrated) energy distribution to the measured transmission function while assuming a fixed angular distribution at a small pitch angle. For the case discussed here, an angular distribution with mean angle $\hat{\theta} = \ang{2}$ and angular spread $\sigma_\theta = \ang{1}$ was used. These values are consistent with particle-tracking simulations (section~\ref{simulations:plate_angles}) and complementary measurements of the angular distribution that were performed at the monitor spectrometer (section~\ref{measurements:transmission}).

The measurements discussed in this section use the analytical transmission model \eqref{eq:transmission_model} with five free parameters: the amplitude and background of the electron signal, as well as the mean, width and shape of the energy distribution.
The statistical uncertainty at each data point is derived from the measured rate fluctuations by computing the median- and $1\sigma$-percentiles of the rate taken at \SI{2}{s} intervals for each data run at a fixed value of $U_\mathrm{start}$ (constant surplus energy). In most cases, the uncertainty determined by this method matches the $\sqrt{N}$ expectation from Poisson statistics. However, the percentile method is believed to be more robust against asymmetric rate fluctuations, and is thus preferred. For the transmission function measurements, an uncertainty of $\pm \SI{60}{meV}$ is assumed for the surplus energy, which is included in the fit as an additional term in the uncertainty of each data point. The value has been estimated from the contributions of the individual power supplies that are used in the setup~\cite{erhard:phd}.

Figure~\ref{fig:linewidth} shows transmission functions that were measured using the UV laser (\SI{266}{nm}) and UV LEDs (\SIrange{272}{302}{nm}) at $U_\mathrm{ana} = \SI{-18.6}{kV}$ and $B_\mathrm{min} = \SI{0.38}{mT}$. The laser measurement was performed twice and produced consistent results in terms of the corresponding energy distribution. All measurements used the $U_{dip} = \SI{2}{kV}$ setting that was determined earlier. The observed transmission functions have been normalized to show a transmission probability with an average background of \num{0} and an average nominal amplitude of \num{1} in the plot. Statistical fluctuations can yield negative amplitudes, as seen in the \SI{302}{nm} measurement. This normalization procedure was also applied in the subsequent measurements.
When the UV-LEDs are used as light sources, the observed width of the transmission function decreases for larger wavelengths. This matches the expectation that the photoelectrons emitted from the photocathode material have a lower energy spread when the wavelength is closer to the work function of the photocathode, according to \eqref{eq:photoeffect}.
The energy distributions that were determined from the observed transmission functions at different wavelengths are shown in the lower panel of fig.~\ref{fig:linewidth}. The reduced energy spread for increasing UV wavelengths is clearly visible. The asymmetric shape of the energy distribution that is observed here is expected from the theoretical model of the photoeffect~\cite{Photoemission1964a}. At lower photon energies, the low-energy fraction of the underlying energy distribution is cut off at $E = \SI{0}{eV}$, which results in a more symmetric shape of the observed distribution.

Table~\ref{tab:linewidth} lists the parameters of the energy distribution, which are derived from the measured transmission functions.
The upper limit of the energy distribution, $E_\mathrm{max}$, is used as an indicator for the achieved energy spread. It corresponds to the energy where the distribution drops to 25\% of its maximum. Using this definition, the range $[0; E_\mathrm{max}]$ includes at least 90\% of the distribution's integral. It is also possible to determine the work function directly from a transmission measurement by relating the measured value of $E_\mathrm{max}$ to the known wavelength $\lambda$. This approach is discussed in section~\ref{measurements:workfunction} below.
The value $\sigma_E$ refers to the width of a symmetric normal distribution, which can be derived from the generalized normal distribution. The transformation to $\sigma_E$ takes into account the asymmetry of the distribution and allows comparing distributions with different asymmetry.

The results indicate that owing to the small angular spread in this setting, the width of the measured transmission function is fully dominated by the energy distribution of the electrons. This is true especially for measurements with zero angle and small wavelengths, where the angular distribution has only a minor effect on the transmission function and the energy spread is comparably large.

\begin{table}[h]
    \centering
    \caption{Measured transmission functions at different wavelengths $\lambda$ and fixed spectrometer voltage $U_\mathrm{spec} = \SI{-18.6}{kV}$ (fig.~\ref{fig:linewidth}). The table shows the upper limit of the energy distribution, $E_\mathrm{max}$, and the energy spread, $\sigma_E$; both values are derived from the fit result. The measurement at \SI{266}{nm} has been performed twice at \SI{-18.6}{kV}.}
    \newcolumntype{L}[1]{>{\raggedright\arraybackslash}p{#1}}
    \newcolumntype{C}[1]{>{\centering\arraybackslash}p{#1}}
    \newcolumntype{R}[1]{>{\raggedleft\arraybackslash}p{#1}}
    \newcommand{\ctab}{\centering\arraybackslash}
    \begin{tabular}{R{10mm}R{16mm}R{16mm}R{12mm}}
    \toprule
        \ctab$\lambda$        &\ctab$E_\mathrm{max}$   &\ctab$\sigma_E$        &\ctab$\chi^2/ndf$      \\
    \midrule
        \SI{266.0}{nm}         &\SI{0.82+-0.02}{eV}    &\SI{0.31+-0.05}{eV}    &\num{1.39}             \\
        \SI{266.0}{nm}         &\SI{0.82+-0.02}{eV}    &\SI{0.28+-0.04}{eV}    &\num{1.40}             \\
        \SI{272.4}{nm}         &\SI{0.74+-0.01}{eV}    &\SI{0.22+-0.02}{eV}    &\num{1.18}             \\
        \SI{282.4}{nm}         &\SI{0.61+-0.02}{eV}    &\SI{0.19+-0.03}{eV}    &\num{1.23}             \\
        \SI{292.4}{nm}         &\SI{0.47+-0.02}{eV}    &\SI{0.14+-0.03}{eV}    &\num{3.38}             \\
        \SI{302.4}{nm}         &\SI{0.33+-0.02}{eV}    &\SI{0.09+-0.07}{eV}    &\num{3.46}             \\
    \bottomrule
    \end{tabular}
    \label{tab:linewidth}
\end{table}

\subsection{Magnetic reflection}
\label{measurements:reflection}

Magnetic reflection occurs when the electron pitch angle reaches \ang{90} and the total kinetic energy is in the transversal component. The pitch angle increases from the source towards $B_\mathrm{max} = \SI{6}{T}$ at the spectrometer entrance magnet as a result of adiabatic transformation \eqref{eq:magnetic_moment}.
Magnetic reflection can be investigated by increasing the plate angle, $\alpha_\mathrm{p}$, until a rate decrease is observed at the detector. To ensure that electrons are reflected only magnetically and not because of an insufficient surplus energy \eqref{eq:transmission_cond}, the measurement is performed at large surplus energies $q \Delta U \ge \SI{10}{eV}$. The rate gradually decreases with increasing $\alpha_\mathrm{p}$ as more electrons are reflected as a result of the angular distribution in the magnet. The rate dependency can be modeled by a symmetric error function, which allows us to investigate the angular distribution at large pitch angles $\theta \rightarrow \ang{90}$. The center position of the error function is referred to as \emph{reflection angle} $\alpha_\mathrm{max}$; it corresponds to the plate angle where 50\% of electrons are reflected.

\begin{figure}[tb]
        \includegraphics[width=\columnwidth]{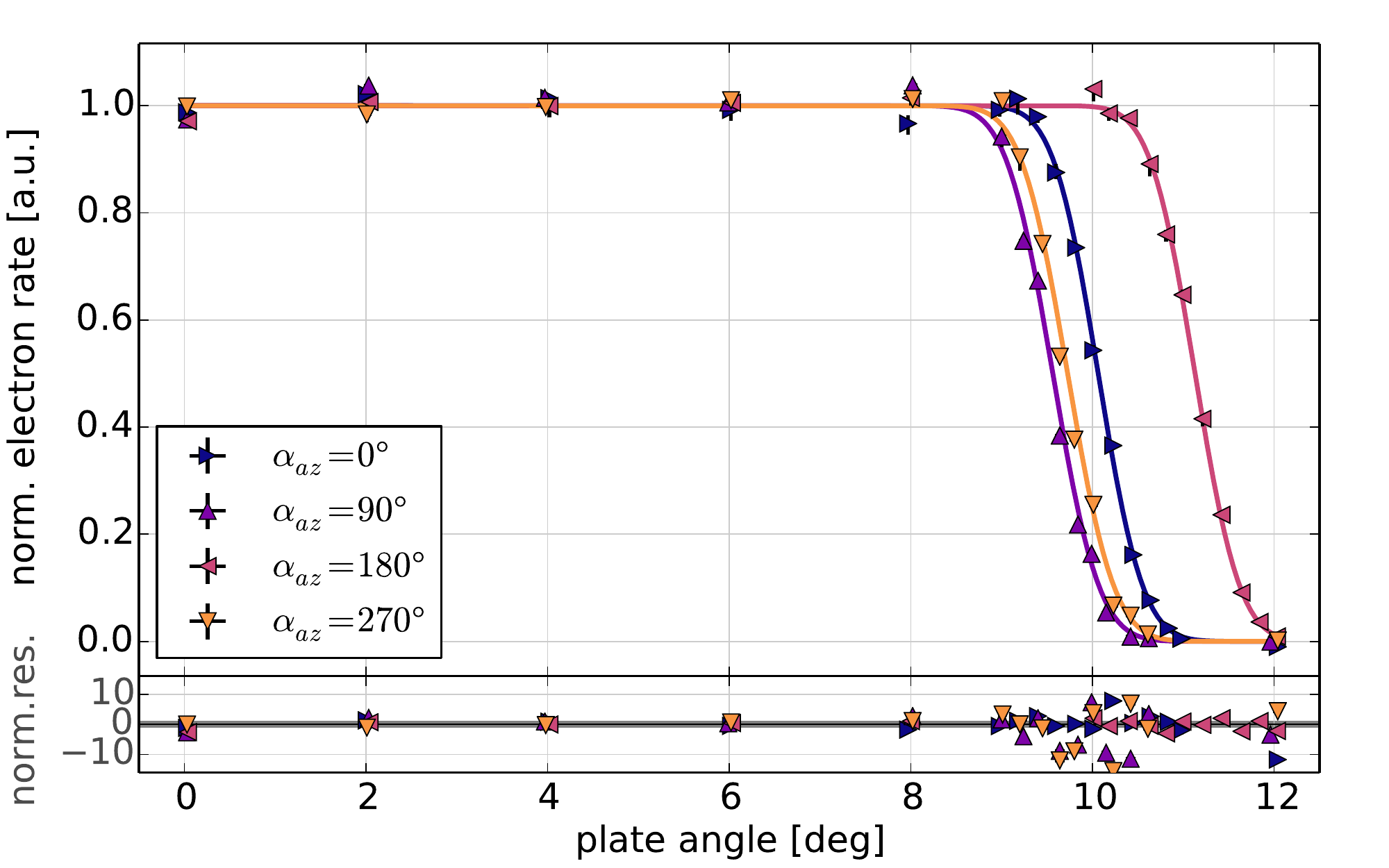}\\%
        \includegraphics[width=\columnwidth]{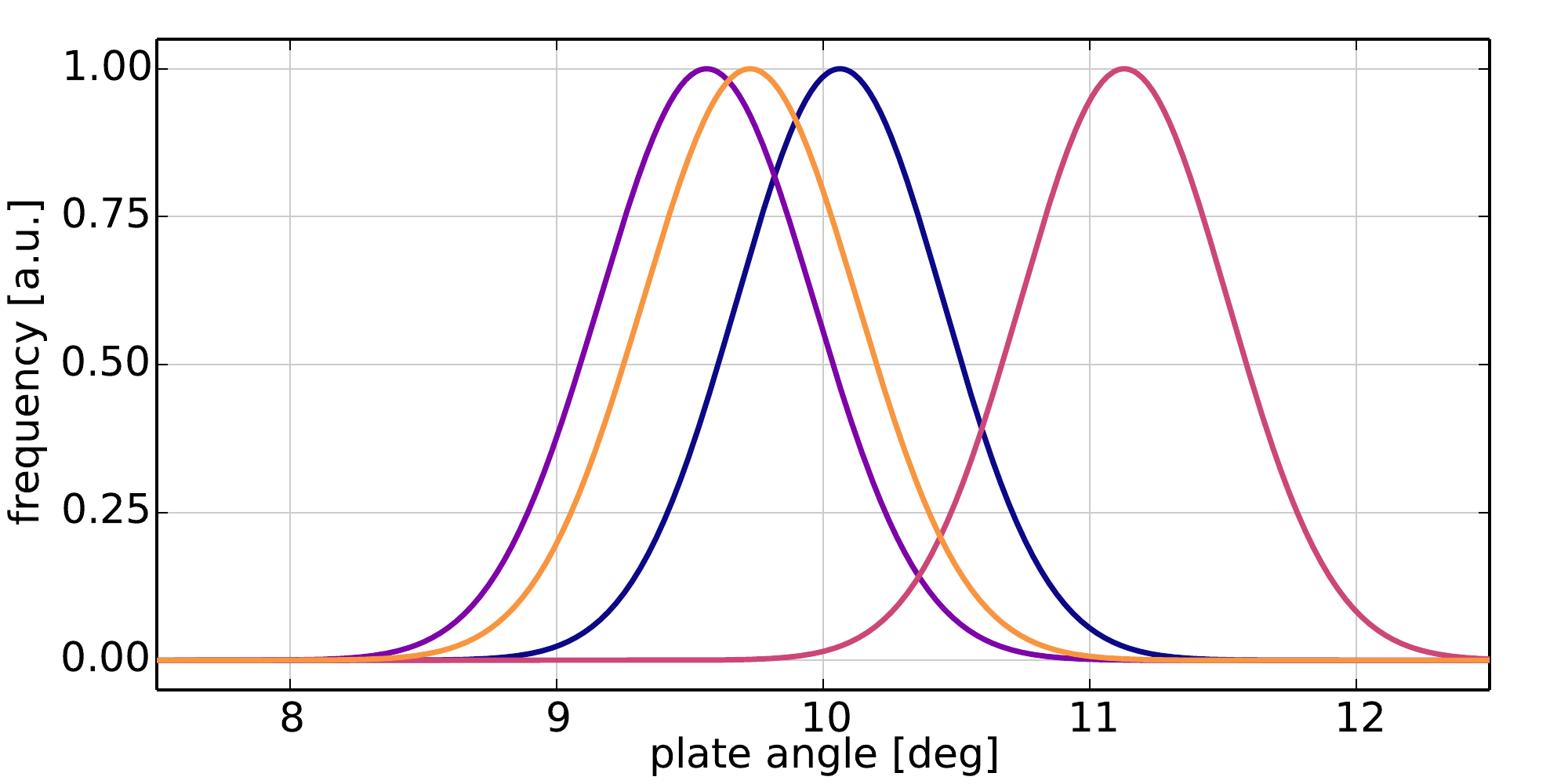}%
        \caption{Measured magnetic reflection curves (top) and angular distributions (bottom) for different azimuthal angles of the source. The spectrometer was operated with $U_\mathrm{spec} = \SI{-18.6}{kV}$ and $B_\mathrm{start}^+ = \SI{27}{mT}$ at the electron emission spot. The given residuals are in absolute units. Large pitch angles are cut off from the transmission function due to magnetic reflection at the spectrometer solenoids ($B_\mathrm{max} = \SI{6}{T}$). By increasing the plate angle $\alpha_\mathrm{p}$, the transmission probability decreases until all electrons are reflected. The point of reflection depends on the ratio of the magnetic fields, $B_\mathrm{start} / B_\mathrm{max}$.
        The angular distributions shown are given in terms of plate angle ($\alpha_\mathrm{p}$); the actual angular spread in terms of pitch angle ($\theta_\mathrm{mag}$) is $\sigma_\theta \approx \ang{16}$.}
    \label{fig:reflection}
\end{figure}

At the monitor spectrometer, this measurement was performed at four different azimuthal directions of the plate angle to investigate possible asymmetries, $\alpha_{az} = \SIlist{0;90;180;270}{\degree}$. The results are shown in fig.~\ref{fig:reflection} (solid lines). The underlying gauss-curves shown at the bottom of the figure allows a better comparison of the angular distributions.

As will be shown later in section~\ref{simulations:plate_angles}, the produced pitch angle $\theta$ increases non-linearly with the plate angle $\alpha_\mathrm{p}$. This results from the adiabatic transformation towards the spectrometer entrance magnet:
\begin{equation}
    \label{eq:adiabatic_transformation}
    \theta  \approx \arcsin\left( \alpha_\mathrm{p} \cdot k \cdot \sqrt{ \frac{B_\mathrm{max}}{B_\mathrm{start}} } \right)
            \approx \arcsin\left( \frac{\alpha_\mathrm{p}}{\alpha_\mathrm{max}} \right)
    \,,
\end{equation}
where $k$ is a scaling factor that depends on the non-adiabatic acceleration of the emitted electrons, and $B_\mathrm{start}$, $B_\mathrm{max}$ are the magnetic fields at the electron source and the spectrometer entrance, respectively.

Table~\ref{tab:reflection} shows the fit results of these measurements. For nominal magnetic field at the electron source ($B_\mathrm{start} = \SI{27}{mT}$), reflection occurs at a plate angle $\alpha_\mathrm{max} \approx \ang{10}$. The width of the angular distribution is consistent over the four measurements, yielding a width of $\sigma_\alpha = \ang{0.40}$ for the underlying Gaussian distribution. The adiabatic transformation \eqref{eq:adiabatic_transformation} converts the value $\sigma_\alpha$ to an effective angular spread $\sigma_\theta$ in the magnet. The conversion employs the constraint that magnetic reflection occurs at $\alpha_\mathrm{p} = \alpha_\mathrm{max}$ with $\theta = \ang{90}$. This yields an average angular spread of $\sigma_\theta = \ang{16.2}$ at the maximal pitch angle of \ang{90}. Note that the angular spread close to magnetic reflection increases because of the non-linearity of \eqref{eq:adiabatic_transformation}, and is significantly lower at smaller pitch angles.

The discrepancy between the measurements in four azimuthal directions can be explained in two ways.
Firstly, particle-tracking simulations indicate that misalignments of the emission spot relative to the plate setup of the electron source result in significant offsets of the produced pitch angles. Such misalignments can result from mechanical imperfections of the setup and are likely the explanation for the observed asymmetry~\cite{behrens:phd}.
Secondly, phase effects can affect the electron acceleration processes in the source. The cyclotron phase of the emitted electrons differs depending on the azimuthal direction into which the electron beam is collimated. This results in slight variations of the produced pitch angle, which depend on the azimuthal plate angle $\alpha_\mathrm{az}$. The asymmetry in vertical direction ($\alpha_\mathrm{az} = \SIlist{0;180}{\degree}$) is further increased by the electric field of the dipole electrode.

The magnetic reflection measurements have been fitted with MINUIT2~\cite{MINUIT1975}, using a normal distribution in integral form (scaled error function) to model the shape of the reflection curve. The uncertainty of the electron rate has been determined like explained above. An uncertainty of \ang{0.05} is assumed for the plate angle and included in the fit; this value corresponds to the uncertainty of the plate angle read-out at the source (section~\ref{design:egun}).

\begin{table}[h]
    \centering
    \caption{Measured magnetic reflection curves at different azimuthal directions $\alpha_\mathrm{az}$ of the plate angle (figure~\ref{fig:reflection}). The table shows the reflection angle $\hat{\alpha} = \alpha_\mathrm{max}$ and the width $\sigma_\alpha$ (in terms of plate angle) of the reflection curve that was determined by the fit. The angular spread $\sigma_\theta$ (in terms of pitch angle) has been computed from the adiabatic transformation \eqref{eq:adiabatic_transformation} with the known reflection angle.}
    \newcolumntype{L}[1]{>{\raggedright\arraybackslash}p{#1}}
    \newcolumntype{C}[1]{>{\centering\arraybackslash}p{#1}}
    \newcolumntype{R}[1]{>{\raggedleft\arraybackslash}p{#1}}
    \newcommand{\ctab}{\centering\arraybackslash}
    \begin{tabular}{R{8mm}R{12mm}R{12mm}R{12mm}R{12mm}}
    \toprule
        \ctab$\alpha_\mathrm{az}$     
                                &\ctab$\alpha_\mathrm{max}$    
                                                        &\ctab$\sigma_\alpha$   &\ctab$\sigma_\theta$    &$\chi^2/ndf$           \\
    \midrule
        \ang{0}                &\SI{10.06(2)}{\degree}  &\SI{0.39(3)}{\degree}  &\SI{16.0(10)}{\degree}  &\num{0.71}             \\
        \ang{180}              &\SI{11.13(3)}{\degree}  &\SI{0.39(3)}{\degree}  &\SI{15.2(12)}{\degree}  &\num{0.88}             \\
        \ang{90}               &\SI{ 9.56(2)}{\degree}  &\SI{0.40(2)}{\degree}  &\SI{16.7(9)}{\degree}   &\num{1.89}             \\
        \ang{270}              &\SI{ 9.73(2)}{\degree}  &\SI{0.40(3)}{\degree}  &\SI{16.6(10)}{\degree}  &\num{0.82}             \\
    \midrule
        \multicolumn{2}{l}{weighted average:}           &\SI{0.396(1)}{\degree} &\SI{16.20(7)}{\degree}  &                       \\
    \bottomrule
    \end{tabular}
    \label{tab:reflection}
\end{table}

\subsection{Angular selectivity}
\label{measurements:transmission}

When the plate angle at the source is increased, a larger pitch angle relative to the magnetic field vector is imprinted on the emitted electrons. The pitch angle in the analyzing plane of the spectrometer is a result of the adiabatic transformation \eqref{eq:magnetic_moment} and thus depends on the produced pitch angle at the source and the magnetic field variation between source and analyzing plane. The increased pitch angle leads to a shift of the measured transmission function to higher surplus energies, as the transversal component of the kinetic energy is larger in this case and needs to be compensated for. Such a shift is only observable if the electron source can produce large-pitch angles with a small angular spread, referred to as \emph{angular selectivity}.
The measured shift between the minimal pitch angle (zero angle $\theta \approx \ang{0}$) and the maximal pitch angle ($\theta = \ang{90}$ in the entrance magnet) allows us to determine the energy resolution \eqref{eq:energy_resolution} of the spectrometer.

\begin{figure}[tb]
    \includegraphics[width=\columnwidth]{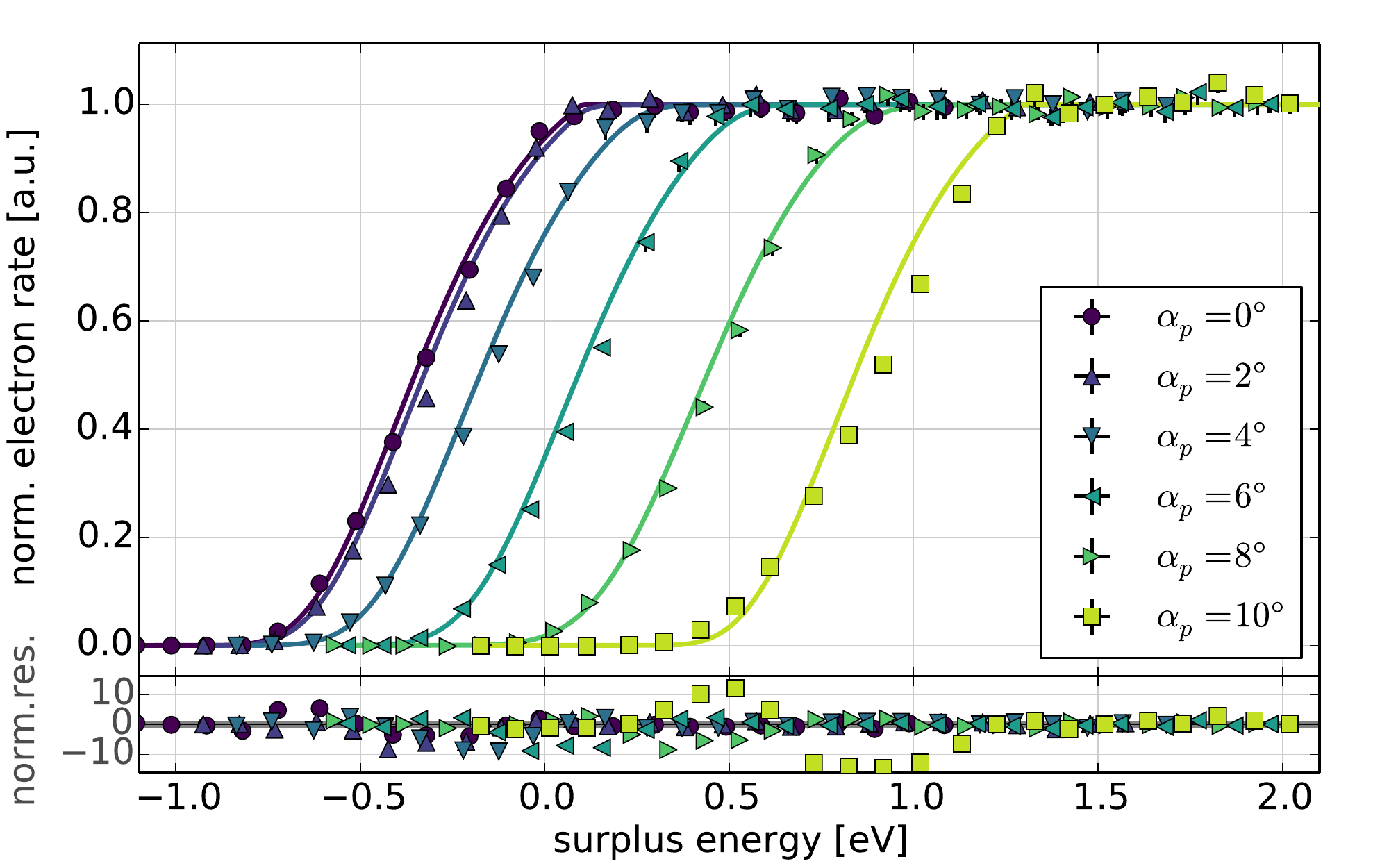}\\%
    \includegraphics[width=\columnwidth]{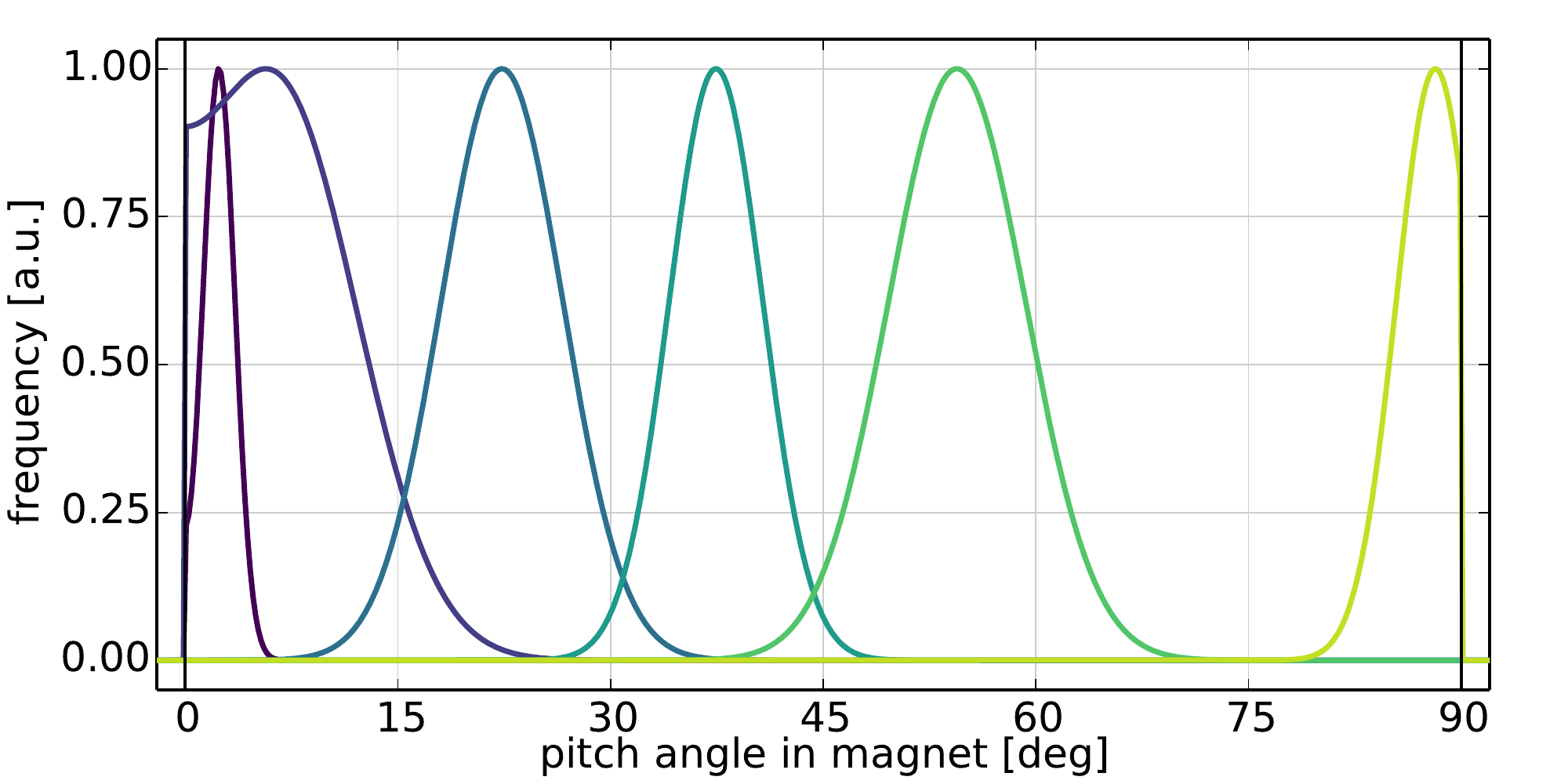}%
    \caption{Measured transmission functions and angular distributions at different plate angles $\alpha_\mathrm{p}$ with $U_\mathrm{ana} = \SI{-18.6}{kV}$ and $B_\mathrm{start}^+ = \SI{27}{mT}$. The given residuals are in absolute units. As expected, a shift towards larger surplus energies is observed for increasing plate angles. Magnetic reflection occurs at $\alpha_\mathrm{p} \approx \ang{10}$, leading to a significant deformation of the transmission function \wrt{} the reference measurement at $\alpha_\mathrm{p} = \ang{0}$.
    The total shift in the transmission functions (\SIrange{0}{10}{\degree}) corresponds to the energy resolution of the spectrometer for an isotropic source. }
    \label{fig:transmission}
\end{figure}

Figure~\ref{fig:transmission} shows measured transmission functions at different plate angles $\alpha_\mathrm{p}$. The zero angle setting $\alpha_\mathrm{p} = \ang{0}$ is used as a reference for the other measurements. The transmission functions are clearly separated and the expected shift to larger surplus energies is observed when increasing the plate angle. Table~\ref{tab:transmission} shows the corresponding parameters of the derived angular distribution.
Magnetic reflection occurs at $\alpha_\mathrm{p} \ge \ang{10}$ (as expected from the magnetic reflection measurement, which yields $\alpha_\mathrm{max} = \ang{10.1}$; cmp. section~\ref{measurements:reflection}). This results in a significantly deformed angular distribution, because reflected electrons are missing from the observed transmission function. Because the fit is based on a reference measurement at $\alpha_\mathrm{p} = \ang{0}$ to obtain the corresponding energy distribution, the deformation affects the fit result and explains the large $\chi^2$ value.

The transmission functions were fitted as explained above. However in this case, free parameters were the amplitude and background of the electron signal and the mean angle and the angular spread. This allows us to determine the produced pitch angle directly from the measurement, while assuming a known energy distribution of the electron source. In this case, a reference measurement at nominal settings ($\alpha_\mathrm{p} = \ang{0}$, $\lambda = \SI{266}{nm}$, $U_\mathrm{ana} = \SI{-18.6}{kV}$) was used for the energy distribution (section~\ref{measurements:linewidth}).
The fit using the analytical model of the transmission function is not very sensitive to the actual shape of the angular distribution for $\theta \rightarrow \ang{0}$ and $\theta \rightarrow \ang{90}$. The angular distribution determined from the measurements at $\alpha_\mathrm{p} = \ang{0}$ and $\alpha_\mathrm{p} = \alpha_\mathrm{max}$ thus yield large uncertainties, and the angular spread is significantly smaller than at intermediate pitch angles. However, the fit results match expectations from an analytical calculation of the pitch angle based on the magnetic reflection limit discussed in section~\ref{measurements:reflection}). The observed pitch angles are also confirmed by simulation results (section~\ref{simulations:plate_angles}).

The total shift between $\alpha_\mathrm{p} = \ang{0}$ and $\alpha_\mathrm{p} = \ang{10}$ corresponds to the maximal difference of pitch angles in the spectrometer entrance magnet, $\theta = \SIrange{0}{90}{\degree}$, and is thus equivalent to the energy resolution of the spectrometer \eqref{eq:energy_resolution}. The observed shift of $\Delta E_\theta = \SI{1.20(6)}{eV}$ corresponds to the expected energy resolution of
\begin{equation}
    \Delta E_\mathrm{ref} = \SI{18.6}{kV} \cdot \frac{\SI{0.38}{mT}}{\SI{6}{T}} = \SI{1.18}{eV}
\end{equation}
for the monitor spectrometer operating at $B_\mathrm{min} = \SI{0.38}{mT}$ at the spectrometer's center and $B_\mathrm{max} = \SI{6}{T}$.

\begin{table}[h]
    \centering
    \caption{Measured transmission functions at different plate angles $\alpha_\mathrm{p}$. The table shows the mean angle $\hat{\theta}$ and the angular spread $\sigma_\theta$ in the spectrometer entrance magnet; the values have been determined by the fit. An expected pitch angle $\hat{\theta}_\mathrm{ana}$ is derived analytically from adiabatic transformation \eqref{eq:magnetic_reflection}. At $\alpha_\mathrm{p} \ge \alpha_\mathrm{max} = \ang{10}$ magnetic reflection is observed, which leads to a significant deformation of the transmission function.}
    \newcolumntype{L}[1]{>{\raggedright\arraybackslash}p{#1}}
    \newcolumntype{C}[1]{>{\centering\arraybackslash}p{#1}}
    \newcolumntype{R}[1]{>{\raggedleft\arraybackslash}p{#1}}
    \newcommand{\ctab}{\centering\arraybackslash}
    \begin{tabular}{R{8mm}R{16mm}R{16mm}R{12mm}R{10mm}}
    \toprule
        \ctab$\alpha_\mathrm{p}$&\ctab$\hat{\theta}$    &\ctab$\sigma_\theta$   &\ctab$\chi^2/ndf$      &\ctab$\hat{\theta}_\mathrm{ana}$ \\
    \midrule
        \ang{0}                 &\SI{ 1.7+-1.3}{\degree}&\SI{2.0+-1.3}{\degree} &\num{1.09}             &\ang{ 2.0}                     \\
        \ang{2}                 &\SI{ 5.7+-3.4}{\degree}&\SI{9.3+-2.6}{\degree} &\num{1.07}             &\ang{13.5}                     \\
        \ang{4}                 &\SI{23.2+-0.3}{\degree}&\SI{5.8+-0.8}{\degree} &\num{1.12}             &\ang{25.4}                     \\
        \ang{6}                 &\SI{38.2+-0.2}{\degree}&\SI{4.3+-0.5}{\degree} &\num{1.31}             &\ang{38.6}                     \\
        \ang{8}                 &\SI{55.2+-0.3}{\degree}&\SI{5.6+-0.4}{\degree} &\num{1.50}             &\ang{54.7}                     \\
        \ang{10}                &\SI{89.3+-0.8}{\degree}&\SI{0.8+-0.7}{\degree} &\num{10.9}             &\ang{85.7}                     \\
    \bottomrule
    \end{tabular}
    \label{tab:transmission}
\end{table}

\subsection{Work function}
\label{measurements:workfunction}

The measured energy spread (section~\ref{measurements:linewidth}) depends on the UV wavelength (photon energy) and the work function of the photocathode material \eqref{eq:photoeffect}. The upper limit of the energy distribution is given by $E_\mathrm{max} = h \nu - \Phi$, and thus the work function can be determined from an energy distribution measurement.
In addition, a direct measurement of the work function is possible by the method conceived by Fowler~\cite{Photoemission1931}. Here the electron yield $I$ is measured at varying UV wavelengths $\lambda$, and the work function $\Phi$ can be determined by fitting the Fowler function to the data,
\begin{align}
    \label{eq:fowler_function}
    I(\mu)      &\propto     T^2 \cdot \xi(\mu)
    \,,
\\
    \xi(\mu)    &=  \begin{cases}
                        e^\mu - \frac{e^{2\mu}}{4} + \frac{e^{3\mu}}{9} + \ldots
                            & (\mu \le 0)
                    \,,
                    \\
                        \frac{\pi^2}{6} + \frac{\mu^2}{2} -
                        \left( e^{-\mu} - \frac{e^{-2\mu}}{4} + \frac{4^{-3\mu}}{9} - \ldots \right)
                            & (\mu > 0)
                    \,,
                    \end{cases}
\end{align}
with $\mu = (h \nu - \Phi) / (k_B T)$. Here $k_B$ is Boltzmann's constant and $T$ the temperature of the photocathode.
In comparison to alternate methods such as using a Kelvin probe~\cite{KelvinProbe1991}, this \emph{in situ} measurement allows us to determine the actual work function of the photocathode under nominal conditions at the experimental site. The determined work function thus can be compared with the measured energy distributions of the electron source.

\begin{figure}[tb]
        \includegraphics[width=\columnwidth]{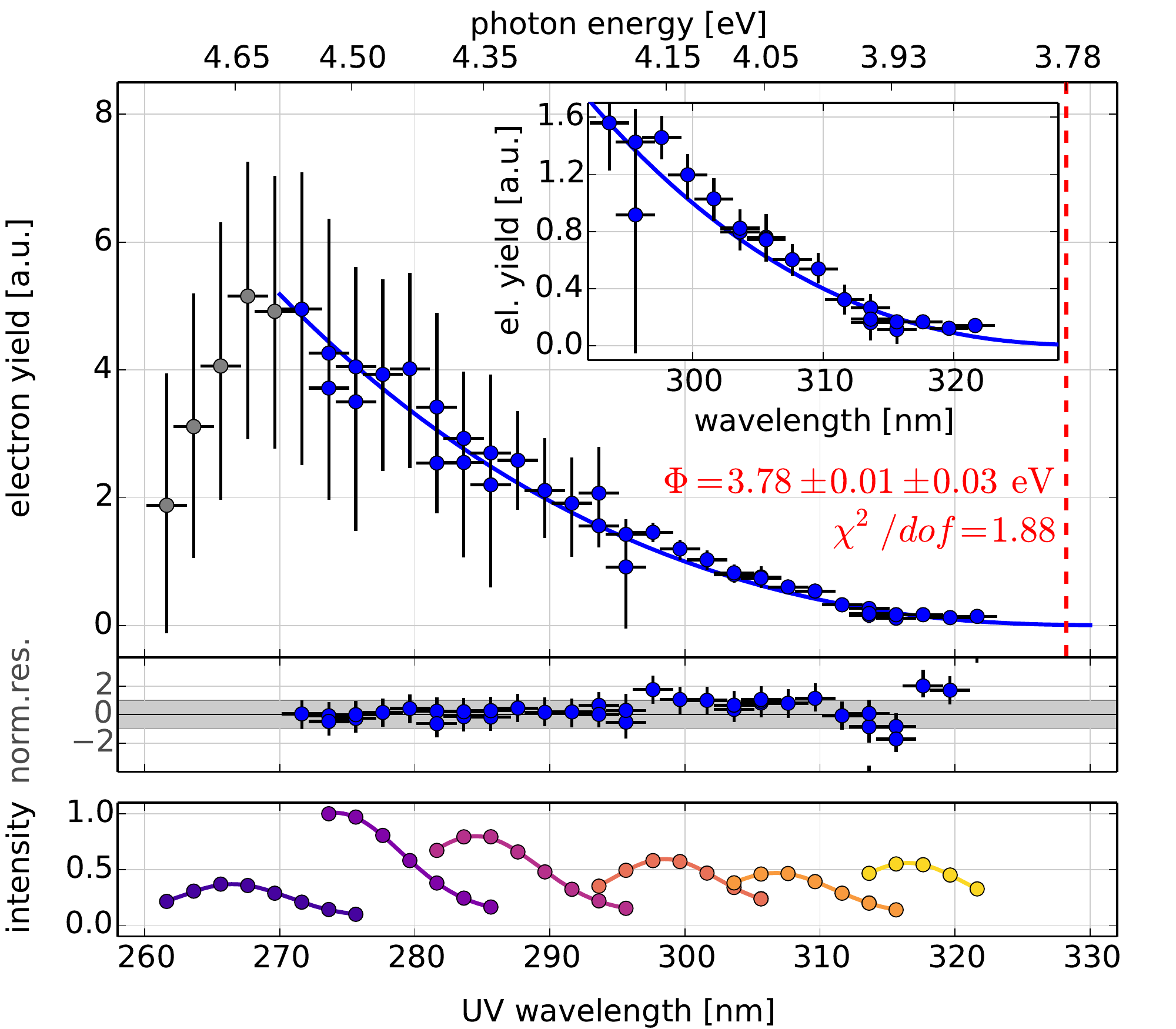}%
        \caption{Measured work function of the photocathode. The work function is determined from the measured electron yield \wrt{} to the UV wavelength, a method proposed by Fowler~\cite{Photoemission1931}. The measurement was performed using six UV LEDs with a monochromator to select wavelengths in the range \SIrange{261}{321}{nm}. The given residuals are in absolute units. The fit results in a work function of $\Phi = \SI{3.78(1)}{eV}$, or a corresponding wavelength of $\lambda_\mathrm{opt} = \SI{326(1)}{nm}$ as indicated by the vertical line. The data points below \SI{270}{nm} were excluded from the fit.
        The scale at the top shows the corresponding photon energy $h \nu$ for the LED peak wavelengths. The bottom plot shows the LED emission profiles with their relative intensity.}
    \label{fig:workfunction}
\end{figure}

Figure~\ref{fig:workfunction} shows the result of a Fowler-type measurement that has been performed at the monitor spectrometer setup. Six different UV LEDs have been used in combination with a monochromator to scan a wavelength range between \SI{261}{nm} and \SI{321}{nm}. The emission profiles of the used LEDs can be approximated by a Gaussian as shown at the bottom of the plot. The achieved electron rate drops significantly when moving away from the LED's peak wavelength, and the lower rate in these cases is compensated by an increased measurement time to reach a similar statistical uncertainty.
For wavelengths $\lambda \gtrsim \SI{270}{nm}$, the observed electron yield matches the expectation from the Fowler equation \eqref{eq:fowler_function}. For smaller wavelengths a deviation from the expected behavior is observed, which gets more emphasized for decreasing wavelengths.
At $\lambda < \SI{270}{nm}$, the electron yield reduces again, which is in contrast to the well-known three-step model of photoemission developed by Berglund and Spicer~\cite{Photoemission1964a,Photoemission1964b}. This observation can be explained by considering effects that become more dominant at higher photon energies~\cite{Photoemission1995}. Close to the work function threshold the photoemission is dominated by excitons, which are created from incident light and move towards the surface where they release their electron (exciton dissociation). At photon energies well above the threshold (here $\gtrsim \SI{4.5}{eV}$), the photoabsorption becomes dominated by electron-excitation into unbound states. These electrons are not emitted from the surface, and the total electron yield is thus reduced at smaller wavelengths.

The fit results in a work function of $\Phi = \SIerr{3.78}{0.03}{0.01}{eV}$ for the gold photocathode with \SI{20}{nm} layer thickness at $T = \SI{300}{K}$. The systematic uncertainty is estimated to \SI{0.03}{eV} from the uncertainty of the wavelength caused by to the filter width of the monochromator ($\SI{0.01}{eV} \hat{=} \SI{1}{nm}$), and the uncertainty of the LED peak wavelength (\SI{2}{nm}). The monochromator was calibrated beforehand using the known wavelength of the UV laser (\SI{266}{nm}).
The determined work function is equivalent to a wavelength of $\lambda_\mathrm{opt} = \SIerr{328.2}{2.3}{0.7}{nm}$. The energy spread of the electron source can be minimized by matching the UV wavelength to this value. Unfortunately, the available LEDs limit the usable wavelength range to about \SI{320}{nm}, as the very low rate at larger wavelengths would require unfeasibly long measurement times. However, even at wavelengths well above \SI{266}{nm}, the optimum for maximal intensity, the resulting energy spread of \SI{0.3}{eV} or less is sufficiently small to determine the transmission properties of the spectrometer (section~\ref{measurements:linewidth}).

The work function determined from this measurement can be compared with the result from investigating the energy distribution. The maximal kinetic energy $E_{e,\mathrm{max}}$ of the photo-electrons emitted by the electron source is given by the relation
\begin{equation}
    \label{eq:workfunction_electron_energy}
    E_{e,\mathrm{max}} = h \nu - \Phi = hc / \lambda - \Phi
    \,,
\end{equation}
with Planck's constant $h$, the speed of light $c$ and the UV wavelength $\lambda$. The work function $\Phi$ can thus be determined from the upper limit of the energy distribution of the photo-electrons, which is given in table~\ref{tab:linewidth} (cmp. figure~\ref{fig:linewidth}). The upper limit shifts to lower values when the UV wavelength is increased and the incident photons have less energy (\ie{} the distribution gets narrower). The resulting work functions from this method should be consistent for measurements performed at different wavelengths. Table~\ref{tab:workfunction} shows the results from using this approach, and compares the determined work function $\Phi^\dagger$ with the work function $\Phi$ yielded by the Fowler-type measurement above.
A combined analysis of the resulting work functions yields $\overline{\Phi^\dagger} = \SI{3.810(1)}{eV}$, using a weighted average that takes into account the uncertainties of $E_{e,\mathrm{max}}$. This result is consistent with the value determined by the Fowler-type measurement. It is thus verified that both methods produce consistent results, and that the determined work function is applicable to the measured transmission functions.

Our determined value for the work function of a gold surface is far below the theoretical expectation of \SIrange{4.2}{5.1}{eV}~\cite{WorkFunction1970,WorkFunction1995}. However, the work function of our photocathode is affected by various effects like the surface roughness, impurities in the material and electric fields at the surface~\cite{WorkFunctionEffects1973,WorkFunctionEffects2005}. These effects are largely eliminated when work functions are determined under ideal laboratory conditions, which makes these results incomparable to our \emph{in situ} measurement. This observation is confirmed by other measurements performed at the KATRIN main spectrometer~\cite{behrens:phd} and in a test setup at WWU M{\"u}nster~\cite{zacher:phd,winzen:diploma}.

\begin{table}[h]
    \centering
    \caption{Work functions determined from measured transmission functions at different wavelengths $\lambda$ (fig.~\ref{fig:linewidth}). The work functions $\Phi^\dagger$ are derived from \eqref{eq:workfunction_electron_energy}, with $E_{e,\mathrm{max}}$ the upper limit of the energy distribution (tab.~\ref{tab:linewidth}) and $hc / \lambda$ the known photon energy. The results are compared with the work function $\Phi = \SI[parse-numbers=false]{3.78(4)}{eV}$ that was determined in a Fowler-type measurement (fig.~\ref{fig:workfunction}).}
    \newcolumntype{L}[1]{>{\raggedright\arraybackslash}p{#1}}
    \newcolumntype{C}[1]{>{\centering\arraybackslash}p{#1}}
    \newcolumntype{R}[1]{>{\raggedleft\arraybackslash}p{#1}}
    \newcommand{\ctab}{\centering\arraybackslash}
    \begin{tabular}{R{10mm}R{16mm}R{16mm}R{16mm}}
    \toprule
        \ctab$\lambda$          &\ctab$hc/\lambda$      &\ctab$\Phi^\dagger$    &\ctab$\Phi^\dagger-\Phi$     \\
    \midrule
        \SI{266.0}{nm}          &\SI{4.66+-0.02}{eV}    &\SI{3.84+-0.04}{eV}    &\SI{ 0.05+-0.07}{eV}   \\
        \SI{266.0}{nm}          &\SI{4.66+-0.02}{eV}    &\SI{3.84+-0.04}{eV}    &\SI{ 0.05+-0.07}{eV}   \\
        \SI{272.4}{nm}          &\SI{4.55+-0.04}{eV}    &\SI{3.81+-0.05}{eV}    &\SI{ 0.02+-0.08}{eV}   \\
        \SI{282.4}{nm}          &\SI{4.39+-0.04}{eV}    &\SI{3.79+-0.05}{eV}    &\SI{ 0.00+-0.08}{eV}   \\
        \SI{292.4}{nm}          &\SI{4.24+-0.03}{eV}    &\SI{3.77+-0.05}{eV}    &\SI{-0.02+-0.08}{eV}   \\
        \SI{302.4}{nm}          &\SI{4.10+-0.03}{eV}    &\SI{3.77+-0.05}{eV}    &\SI{-0.02+-0.08}{eV}   \\
    \midrule
        \multicolumn{2}{l}{weighted average:}           &\SI{3.810+-0.001}{eV}  &\SI{ 0.03+-0.03}{eV}   \\
    \bottomrule
    \end{tabular}
    \label{tab:workfunction}
\end{table}

\section{Simulations}
\label{simulations}

The particle-tracking software \emph{Kassiopeia} was developed as a joint effort from members of the KATRIN collaboration to simulate trajectories of charged particles such as electrons or ions in complex electromagnetic fields with very high precision~\cite{Kassiopeia2016}. Kassiopeia is embedded in the so-called KASPER framework, the overall KATRIN software package.
The software is used to study the transmission properties of the KATRIN spectrometers and to investigate background processes, among other simulation tasks. For the development of the electron source presented in this paper, Kassiopeia simulations provided substantial input for optimizations of the existing design. Detailed simulations were performed to investigate the electron acceleration processes within the source and to understand how the well-defined pitch angles are produced.

\subsection{Implementation into Kassiopeia}
\label{simulations:setup}

Kassiopeia performs tracking of charged particles in electromagnetic fields based upon a given simulation geometry. Electric fields are computed by the \emph{boundary element method} (BEM) from a set of charge densities at the electrode surfaces. The charge densities are pre-computed from the given electrode potentials with the iterative \emph{Robin Hood} method~\cite{RobinHood2012}. For axially symmetric electric fields, an approximation method known as \emph{zonal harmonic expansion} can be used to speed up the field computations with negligible loss of accuracy~\cite{Elcd2011}. To accurately model the electron source with all relevant components (\eg{} the half-shell dipole electrode) it is necessary to use geometric shapes that break axial symmetry, thus no such approximation can be used. KEMField supports OpenGL-based graphics processing unit (GPU) acceleration, a feature that was utilized to considerably reduce the required computation time of such complex geometric structures.
Magnetic fields are computed from a given set of coil geometries (solenoids and air coils) via elliptic integration; it is possible to apply zonal harmonic expansions here as well~\cite{Magfield2011}. The simulations of the electron source use a detailed model of the magnet system at the monitor spectrometer.
The particle-tracking in Kassiopeia is carried out by discretizing the trajectory into a finite number of steps. At each step the electromagnetic fields $\vec{E}(\vec{x}),\vec{B}(\vec{x})$ are evaluated and the equation of motion is solved by integration~\cite{furse:phd,groh:phd,Kassiopeia2016}, after which the particle propagates to the next step. For charged particles, the Lorentz force \eqref{eq:lorentz} defines the equation of motion.

\begin{figure}[tb]
    \includegraphics[width=\columnwidth]{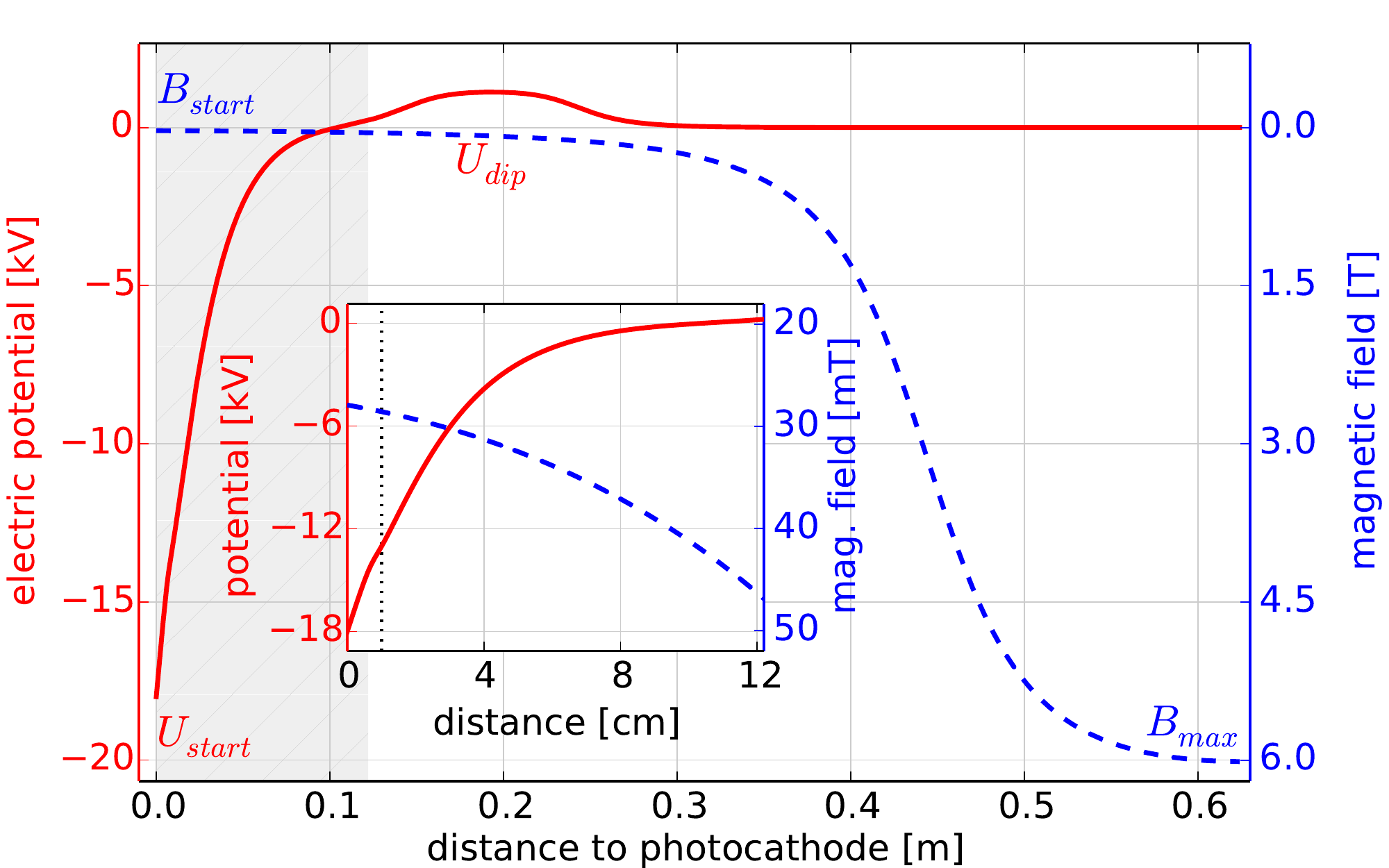}%
    \caption{Simulated electric potential (solid red, left axis) and magnetic field (dashed blue, right axis) at the monitor spectrometer setup \wrt{} to the position of the photocathode. The electrons are emitted at $U_\mathrm{start} = \SI{-18.6}{kV}$ and accelerated towards ground potential. Trapped electrons are removed by a dipole electrode with $U_\mathrm{dip} = \SI{+2}{kV}$. The spectrometer magnet with $B_\mathrm{max} = \SI{6}{T}$ in combination with an additional air coil at the electron source also defines the initial magnetic field $B_\mathrm{start}^+ = \SI{27}{mT}$. The length of the source cage corresponds to the shaded region, which is enlarged in the inset. An acceleration potential $U_\mathrm{acc} = \SI{+5}{kV}$ is applied at the front plate (indicated by a dotted line in the inset).}
    \label{fig:sim_fields}
\end{figure}

The electrode geometry of the electron source was implemented in Kassiopeia based on CAD drawings of the electron source design.
The position of the electrodes \wrt{} the spectrometer setup was determined from measurements at the experimental site and from comparisons of simulated with measured magnetic fields (section~\ref{measurements:setup}).
Figure~\ref{fig:sim_fields} shows the simulated magnetic field and electric potential between the photocathode of the electron source and the entrance magnet of the monitor spectrometer.

\subsection{Energy and angular distributions}
\label{simulations:energy_angle}

The simulations allow us to investigate the electron acceleration mechanisms inside the source. An important question is the effect of the electromagnetic fields on the energy and angular distributions achieved. Electrons were started from the emission spot on the back plate (radius \SI{100}{\micro m}, according to the dimensions of the optical fiber in the experimental setup), where the starting voltage $U_\mathrm{start} \approx \SI{-18.6}{kV}$ is applied. The initial energy is normal-distributed in the range \SIrange{0}{0.6}{eV} ($\mu = \sigma = \SI{0.2}{eV}$). The initial polar angle \wrt{} the back plate follows a $\cos\theta$-distribution in the range \SIrange{0}{90}{\degree}. The parameters of the energy distribution were chosen according to measurement results, which yield an energy spread of up to \SI{0.3}{eV} (section~\ref{measurements:linewidth}), while the angular distribution matches the results from \cite{Photoemission2002}.

In the simulations presented here, \num{1000} electrons were created at the back plate for each setting. The electrons were tracked up to the spectrometer entrance magnet in order to determine the energy and angular distributions. Both distributions are key parameters for the analysis of transmission function measurements.

\begin{figure}[tb]
    \includegraphics[width=\columnwidth]{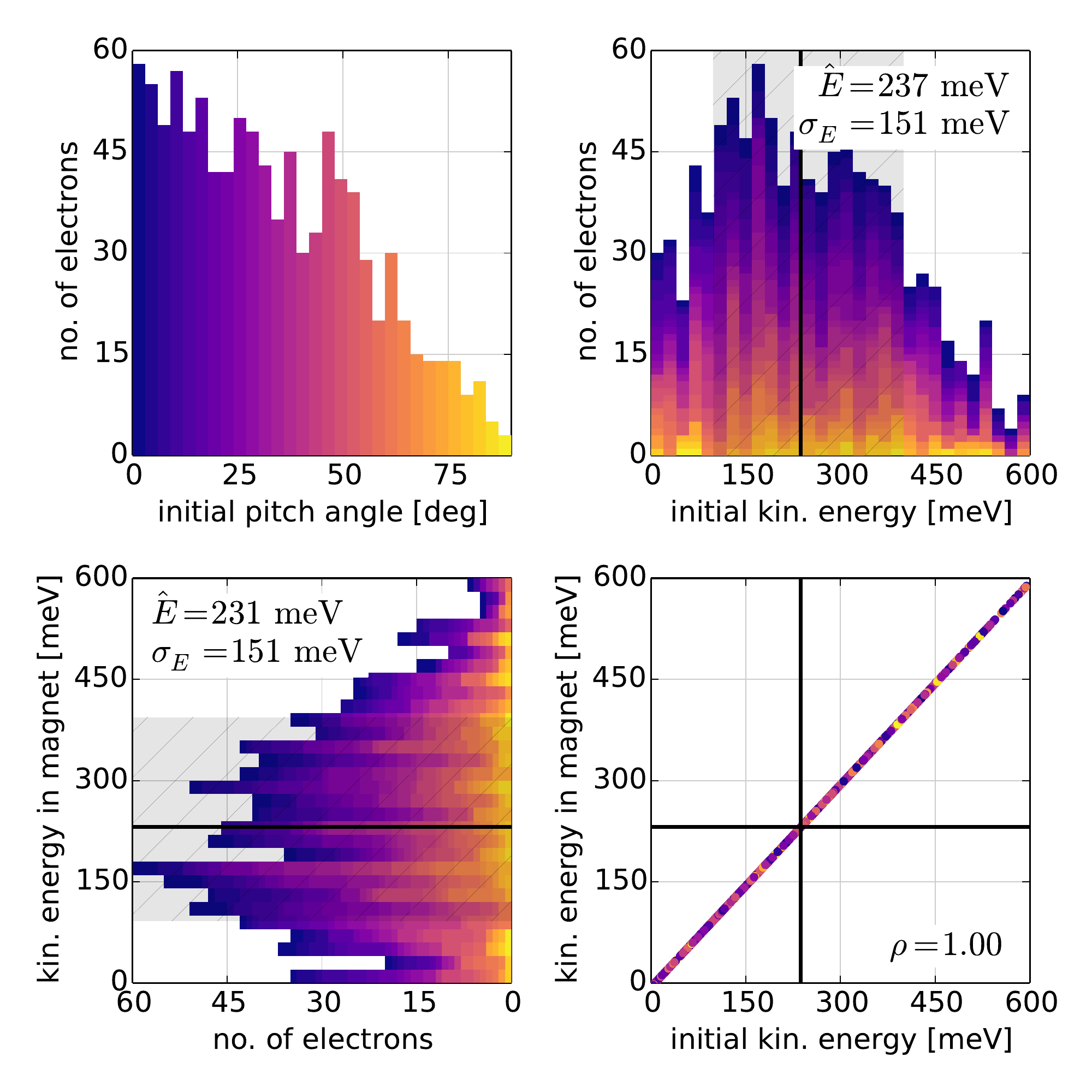}%
    \caption{Simulated energy distributions of \num{1000} electrons at $U_\mathrm{start} = \SI{-18598.7}{V}$. The plot shows the initial energy distribution, $E_e$, on the upper right; the shifted energy distribution in the spectrometer entrance magnet, $E = E_\mathrm{kin} + q U_\mathrm{start}$, on the bottom left; and the correlation between the two distributions on the bottom right. The initial energy $E_e$ corresponds to the start of the electron trajectory at \SI{10}{nm} distance from the photocathode. The distributions are colored by the initial pitch angle, which is $\cos\theta$-distributed as shown on the upper left. The median and the $1\sigma$-width of each distribution is indicated by the black lines and the shaded areas, respectively.}
    \label{fig:sim_distribution_energy}
\end{figure}

\begin{figure}[tb]
    \includegraphics[width=\columnwidth]{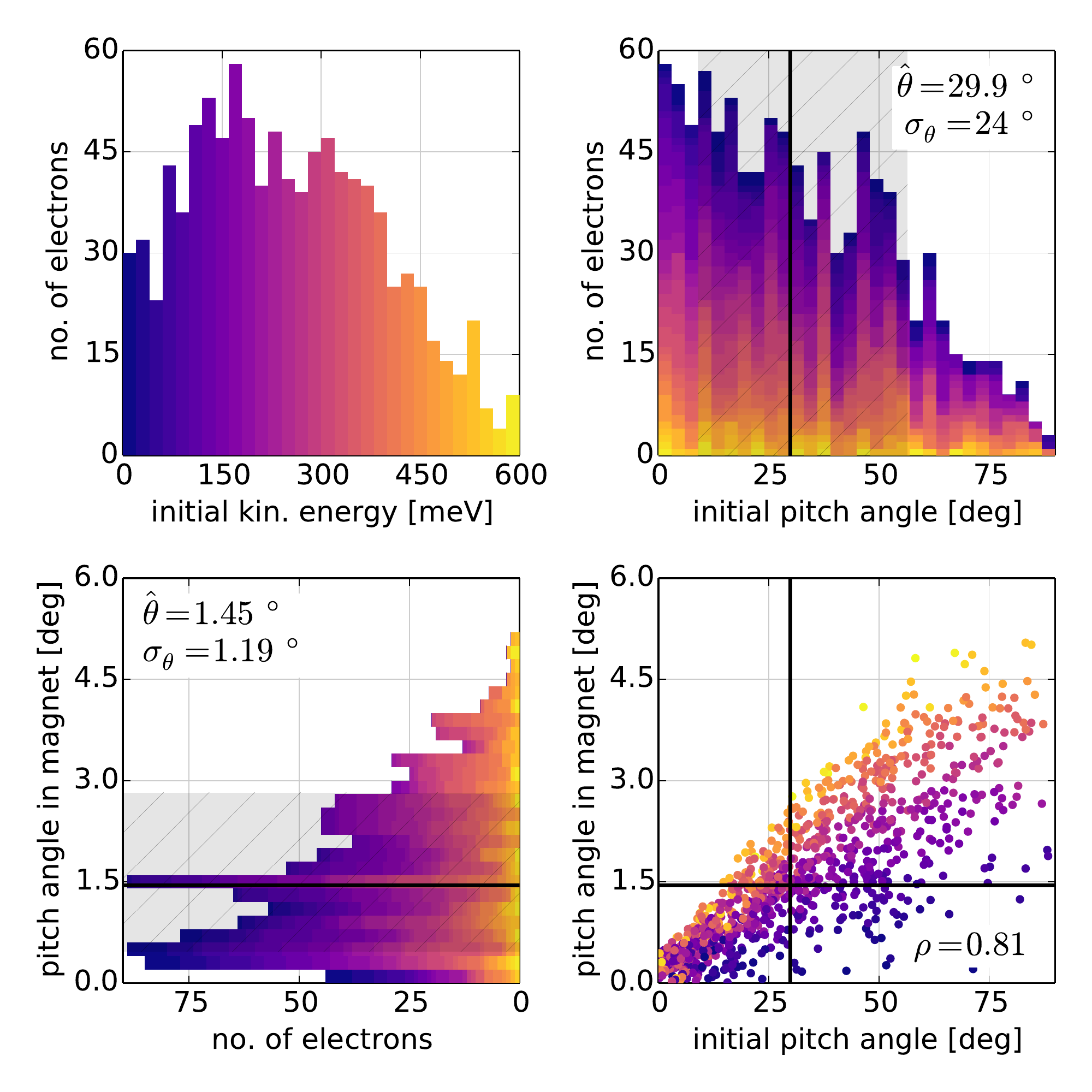}%
    \caption{Simulated angular distributions of \num{1000} electrons at $U_\mathrm{start} = \SI{-18598.7}{V}$. The plot uses the same approach as fig.~\ref{fig:sim_distribution_energy}, but shows the correlation between the initial pitch angle, $\theta_0$, and the produced pitch angle in the spectrometer entrance magnet, $\theta$. Here the distributions are colored by the initial energy, which is normal-distributed as shown in the upper left panel.}
    \label{fig:sim_distribution_angles}
\end{figure}

Figure~\ref{fig:sim_distribution_energy} shows the correlation between the initial and the final energy distributions \wrt{} the initial pitch angle. The simulations used the zero-angle setting ($\alpha_\mathrm{p} = \ang{0}$) at $U_\mathrm{acc} = \SI{5}{kV}$ and $U_\mathrm{dip} = \SI{3}{kV}$\footnote{Note that the simulations use a \SI{3}{kV} dipole setting instead of the \SI{2}{kV} setting used in the measurements discussed above, because they are intended to be comparable with later measurements carried out at the KATRIN main spectrometer.}.
The distributions are characterized by their median and the $1\sigma$-width, which are both computed using quantiles. The energy distribution in the magnet yields a median energy of $\hat{E} = \SI{0.24}{eV}$ with an asymmetric width of $\sigma_E^- = \SI{0.14}{eV}$ and $\sigma_E^+ = \SI{0.16}{eV}$, which is equivalent to the initial energy distribution. The observed asymmetry results from excluding negative energies from the underlying normal distribution ($E \ge \SI{0}{eV}$).
It can be seen that the energy distribution is completely unaffected by the acceleration processes inside the electron source. The resulting distribution is shifted to larger energies by the electrostatic acceleration, $E = E_0 + q U_\mathrm{start}$, but consistent in width and shape. Further simulations showed that this is also true for different values of $U_\mathrm{start} \neq \SI{-18.6}{kV}$ and non-zero plate angles $\alpha_\mathrm{p} > \ang{0}$.
The measured energy distribution in the magnet (section~\ref{measurements:linewidth}) is therefore fully equivalent to the initial energy distribution at the photocathode. This allows investigating the energy spread of the generated electrons by transmission function measurements, and to determine the work function of the photocathode according to \eqref{eq:photoeffect} (section~\ref{measurements:workfunction}).

Figure~\ref{fig:sim_distribution_angles} shows results of the same simulation, but here the correlation between the initial and final angular distributions is investigated \wrt{} the initial kinetic energy.
The electron pitch angle is changed by the non-adiabatic acceleration and the subsequent adiabatic transport to the entrance magnet. While the initial pitch angles follow a $\cos\theta$-distribution with angles up to \ang{90}, the pitch angles in the magnet are narrowly distributed. As above, the distributions were analyzed by their median and width. The distribution in the magnet has a median pitch angle of $\hat{\theta} = \ang{1.5}$ with an asymmetric width of $\sigma_\theta^- = \ang{1.0}$ and $\sigma_\theta^+ = \ang{1.4}$. Here the asymmetry is caused by the fact that the pitch angle is limited to the range \SIrange{0}{90}{\degree} by definition. Whenever the pitch angle would assume negative values resulting from adiabatic transformation, it is instead mirrored to a positive value. The observed distribution is therefore ``wrapped'' into the positive regime at $\theta = \ang{0}$, and thus becomes asymmetric when this effect occurs.
As indicated by the coloring in the figure, the kinetic energy of the electrons influences the produced angular distribution as well. Electrons with higher kinetic energies contribute more to the observed angular spread than low-energetic electrons. The same effect is also observed at larger plate angles $\alpha_\mathrm{p} > \ang{0}$. This is explained by the efficiency of the non-adiabatic acceleration in the plate setup of the electron source, which is responsible for imprinting a well-defined pitch angle on the electrons.
According to the Lorentz equation \eqref{eq:lorentz}, the electrostatic acceleration becomes less effective as the electron energy increases (cmp. section~\ref{design:principle}). Low-energetic electrons are therefore more strongly collimated, while for electrons with higher initial energies the observed angular spread increases. It is thus possible to further reduce the angular spread by tuning the electron source to produce a small energy spread, which can be achieved by matching the UV wavelength to the photocathode work function.

\subsection{Electron acceleration and transport}
\label{simulations:trajectories}

\begin{figure}[tb]
    \includegraphics[width=\columnwidth]{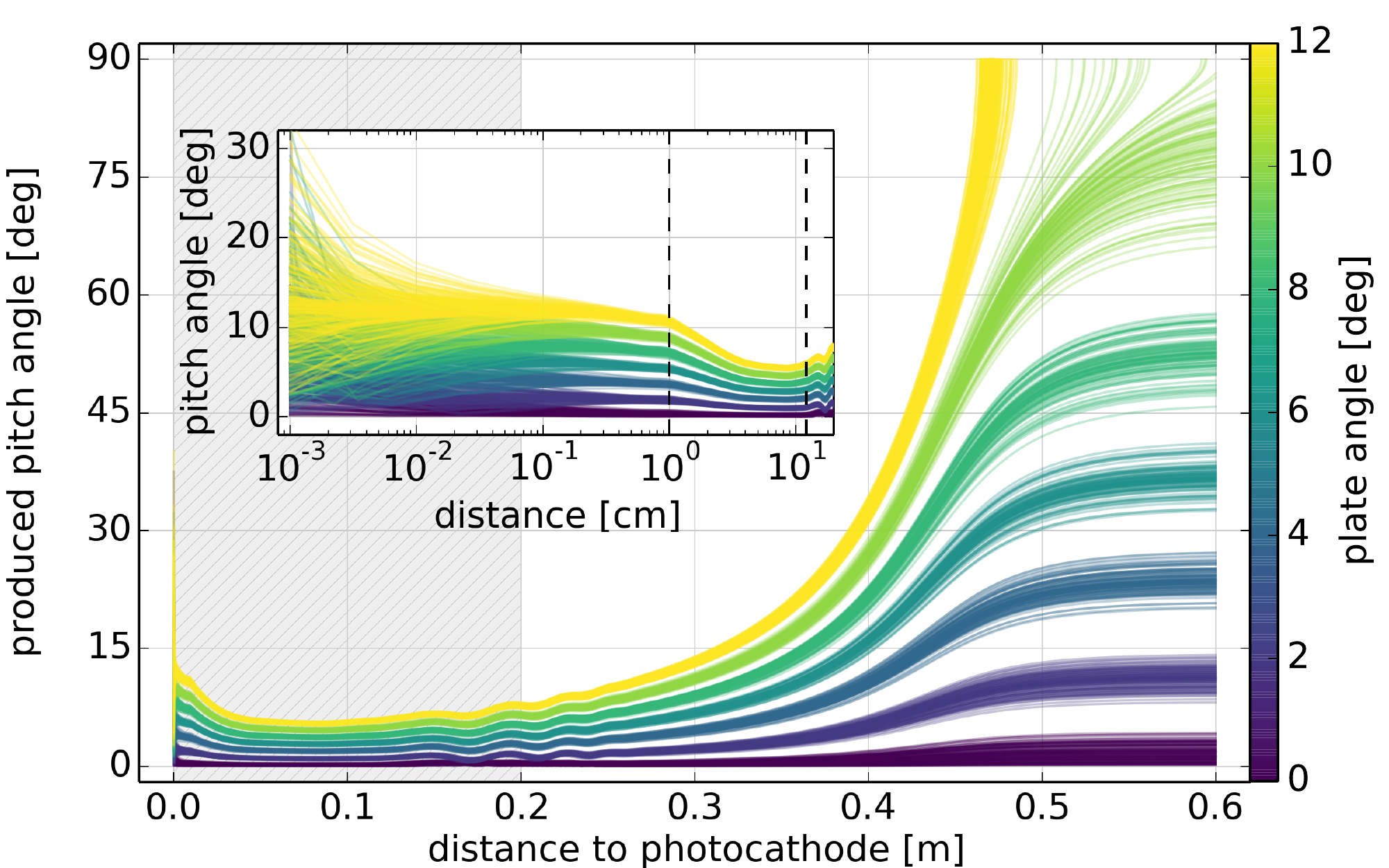}%
    \caption{Simulated pitch angle transformation between the photocathode and the spectrometer entrance magnet for different plate angles $\alpha_\mathrm{p} = \SIrange{0}{10}{\degree}$. The plot shows the evolution of the pitch angle $\theta$ as a function of the distance $d$ from the photocathode. The initial pitch angles are $\cos\theta$-distributed and collimated into a narrow distribution by the strong electric acceleration field at $d \lesssim \SI{1}{cm}$. For $\alpha_\mathrm{p} \ge \ang{10}$, electrons are magnetically reflected before reaching the magnet.
    The inset uses a logarithmic axis to focus on the conditions close to the photocathode where the non-adiabatic acceleration takes place; this is indicated by the shaded region in the main plot. The dashed lines mark the position of the front plate ($d = \SI{1}{cm}$) and the end of the electron source cage ($d = \SI{12.2}{cm}$).}
    \label{fig:sim_angle_trajectories}
\end{figure}

The performance of the electron selectivity in the electron source can be assessed by an investigation of the pitch angle transformation for $\alpha_\mathrm{p} > \ang{0}$. Figure~\ref{fig:sim_angle_trajectories} shows the evolution of the pitch angles along the electron trajectory between the photocathode and the spectrometer entrance magnet. The produced pitch angle along the electron trajectory depends on the plate angle $\alpha_\mathrm{p}$, as indicated by the color scheme. The initial pitch angles are quickly collimated into a narrow distribution. Already at a distance $d \lesssim \SI{1}{mm}$ from the photocathode, the electron beam reaches an angular spread of less than $\ang{0.5}$ for any given setting of $\alpha_\mathrm{p}$.

Electrons that pass the front plate are further accelerated to their full kinetic energy $E = q U_\mathrm{start}$ inside the source cage and transported adiabatically towards the spectrometer magnet. Because transmitted electrons pass the dipole electrode only once, the electric dipole field has no significant influence on the pitch angle transformation. However, the stray electric field of the dipole electrode affects the electron acceleration process itself: because of the asymmetric dipole field, a vertical electric field gradient is generated inside the source cage. Depending on the cyclotron phase of the electrons (and thus, depending on the azimuthal plate angle $\alpha_\mathrm{az}$) the electrons are accelerated differently and the pitch angle changes accordingly. Simulations show that dipole voltages $U_\mathrm{dip} = \SIrange{0}{4}{kV}$ lead to deviations of the pitch angle up to \ang{2}, an observation also made by corresponding measurements. The deviations can be corrected by an empirical determination of the zero angle (cmp. section~\ref{measurements}).
The pitch angle increases towards $B_\mathrm{max}$ as a result of adiabatic transformation. When the pitch angle exceeds $\theta_\mathrm{max} = \ang{90}$, electrons are magnetically reflected. These electrons can get stored between the photocathode and the entrance magnet and need to be removed by the dipole electrode to avoid a possible Penning discharge.

\subsection{Production of well-defined pitch angles}
\label{simulations:plate_angles}

\begin{figure}[tb]
    \includegraphics[width=\columnwidth]{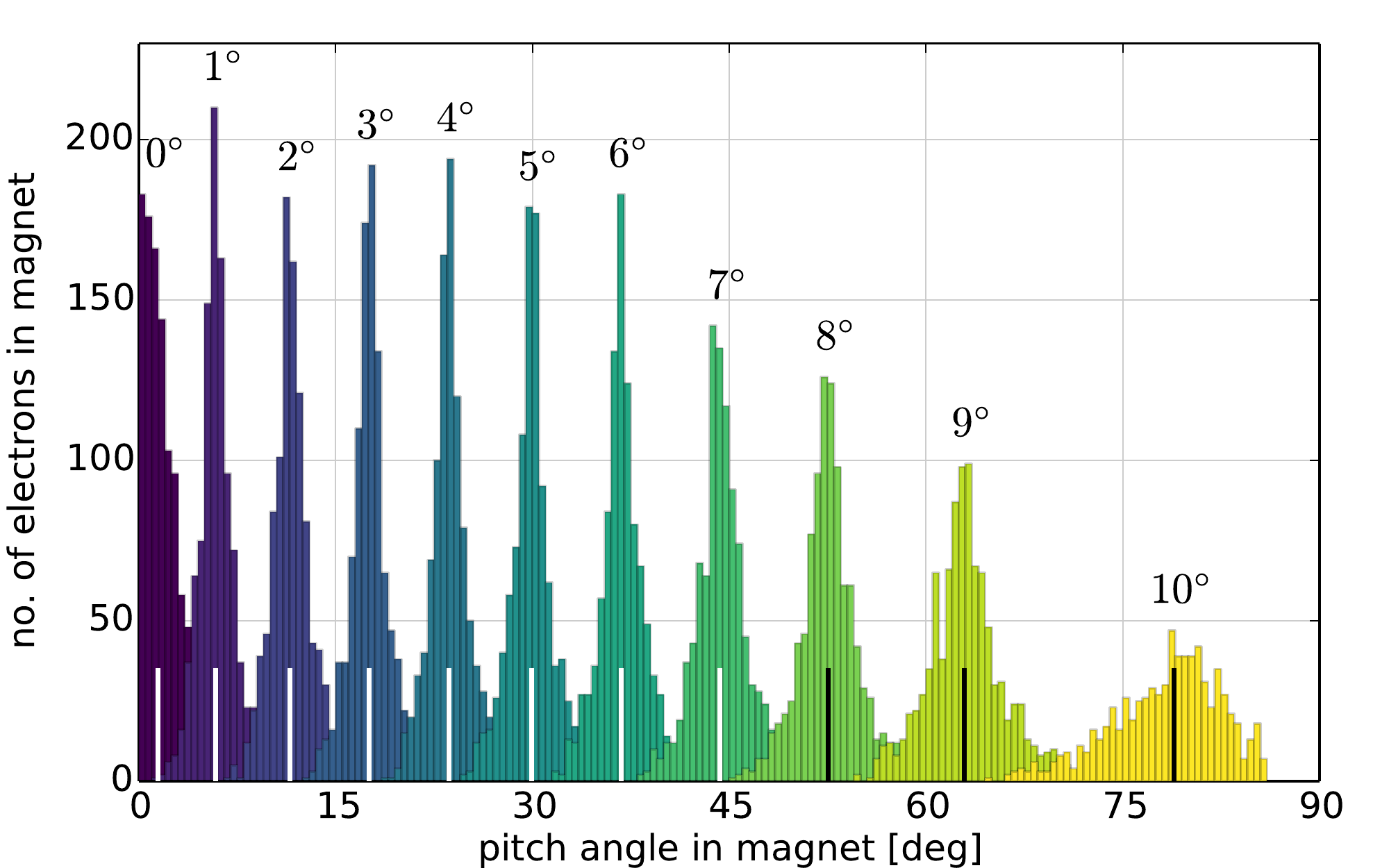}%
    \caption{Simulated pitch angle distributions at different plate angles $\alpha_\mathrm{p} = \SIrange{0}{10}{\degree}$. The plot shows the produced pitch angles in the spectrometer entrance magnet, using same data set as in fig.~\ref{fig:sim_angle_trajectories}. The vertical lines at the bottom mark the median pitch angle of each distribution. For $\theta \rightarrow \ang{90}$ the distribution becomes asymmetric because of magnetic reflection.}
    \label{fig:sim_plate_distributions}
\end{figure}

Figure~\ref{fig:sim_plate_distributions} shows the simulated angular distributions in the spectrometer entrance magnet that are produced by the angular-selective electron source.
When $\alpha_\mathrm{p}$ is increased, the angular distribution is shifted towards larger pitch angles, while the angular spread and shape is not affected; for plate angles $\alpha_\mathrm{p} \gtrsim \ang{8}$, a broadening is observed as $\theta$ approaches the \ang{90} limit. In case of the $\alpha_\mathrm{p} = \ang{10}$ setting, only a low number of electrons reaches the magnet, as the majority is magnetically reflected. At $\alpha_\mathrm{p} = \ang{0}$, the observed distribution is asymmetric because the pitch angle cannot reach negative values; hence, the negative part of the distribution is mirrored at $\theta = \ang{0}$.

\begin{figure}[tb]
    \includegraphics[width=\columnwidth]{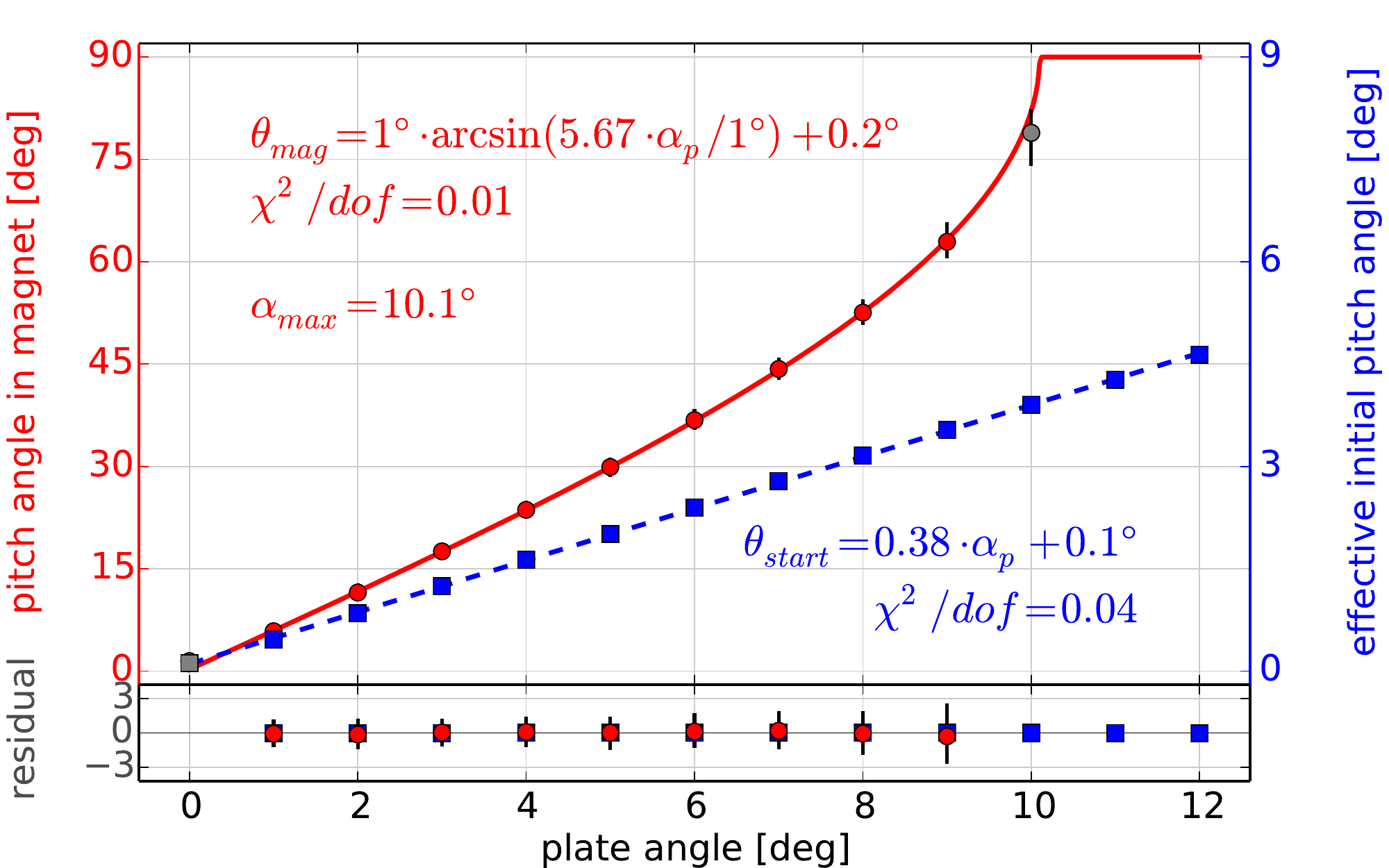}%
    \caption{Simulated pitch angles at different plate angles $\alpha_\mathrm{p}$. The plot shows the produced pitch angle in the spectrometer entrance magnet and the effective initial pitch angle at the photocathode, which is derived according to \eqref{eq:egun_angle_transformation}. The pitch angle in the magnet shows the expected $\arcsin$-dependency on the plate angle, while the initial pitch angle depends linearly on the plate angle.
    Electrons are magnetically reflected at $\alpha_\mathrm{max} = \ang{10.1}$. The data point at $\alpha_\mathrm{p} = \ang{10}$ is affected by partial magnetic reflection and shifted to lower values; it is therefore excluded from the fit.}
    \label{fig:sim_plate_angles}
\end{figure}

The dependency of $\theta$ in the entrance magnet on the plate angle $\alpha_\mathrm{p}$ can be described by the adiabatic transformation \eqref{eq:magnetic_moment} of an electron that propagates towards the spectrometer entrance magnet where a magnetic field $B_\mathrm{max}$ is achieved. The electron starts from the photocathode with an effective starting angle $\theta_\mathrm{start}^*$ (cmp. table~\ref{tab:sim_plate_distributions}) at the initial magnetic field $B_\mathrm{start}$:
\begin{equation}
    \theta = \arcsin\left( \sqrt{ \sin^2 \theta_\mathrm{start}^* \cdot \frac{B_\mathrm{max}}{B_\mathrm{start}} } \right)
    \,.
\end{equation}
One can assume a strictly linear dependency of the small effective starting angle $\theta_\mathrm{start}^*$ and the mechanical plate angle $\alpha_\mathrm{p}$, $\theta_\mathrm{start}^* = f(\alpha_\mathrm{p}) \approx k \cdot \alpha_\mathrm{p}$, which is indicated in fig.~\ref{fig:sim_plate_angles}. A relation between $\alpha_\mathrm{p}$ and the produced pitch angle $\theta$ can be derived by employing the approximation $\sin^2 x \approx x^2$ for small $x$:
\begin{equation}
    \label{eq:egun_angle_transformation}
    \theta = \arcsin\left( \alpha_\mathrm{p} \cdot k \cdot \sqrt{ \frac{B_\mathrm{max}}{B_\mathrm{start}} } \right)
    \,.
\end{equation}
Because the transformation is fully adiabatic, it depends only on the ratio of magnetic fields, $B_\mathrm{start}$ and $B_\mathrm{max}$. This follows from \eqref{eq:magnetic_moment} and \eqref{eq:energy_trans_long} with the kinetic energy $E_0 = q U_\mathrm{start} = \const$ The factor $k$ then describes the effect of the non-adiabatic acceleration in the electron source, which produces the effective initial pitch angle $\theta_\mathrm{start}^*$ at the end of the grounded source cage.

Table~\ref{tab:sim_plate_distributions} lists the corresponding pitch angles and angular spreads, which correspond to the median and the $1\sigma$-width of the angular distributions. Again, the values were computed using percentiles. It should be noted that the pitch angle at $\alpha_\mathrm{p} = \ang{0}$ is systematically larger because of the asymmetric shape of the distribution, which shifts the median to larger values. Similarly, the median at $\alpha_\mathrm{p} = \ang{10}$ is systematically smaller due to the deformation of the angular distribution, which is caused by magnetic reflection.
The angular spread is comparable over a wide range of plate angles with $\sigma_\theta \approx \ang{1.5}$. The spread becomes significantly larger for $\theta \rightarrow \ang{90}$ as a result of adiabatic transformation \eqref{eq:egun_angle_transformation}. The simulated pitch angles and the angular spread are in good agreement with the corresponding measurements (section~\ref{measurements:transmission}).
Table~\ref{tab:sim_plate_distributions} also lists the measured pitch angles $\theta_\mathrm{meas}$ (cmp. tab.~\ref{tab:transmission}), and shows that both results are typically in agreement. An effective starting angle $\theta_\mathrm{start}^*$ has been computed via \eqref{eq:egun_angle_transformation}, showing a strictly linear relation to the plate angle $\alpha_\mathrm{p}$.

\begin{table}[h]
    \centering
    \caption{Simulated pitch angles in the spectrometer entrance magnet and derived effective starting angles. The table shows the median pitch angle, $\hat{\theta}$, and the angular spread, $\sigma_\theta$, in the entrance magnet for different plate angles $\alpha_\mathrm{p}$. The simulation results are compared with the pitch angle determined from corresponding measurements, $\hat{\theta}_\mathrm{meas}$.
    An effective initial pitch angle at the photocathode, $\theta_\mathrm{start}^*$, can be computed from $\hat{\theta}$ using the adiabatic transformation \eqref{eq:egun_angle_transformation} and the known magnetic fields at the setup ($B_\mathrm{start} = \SI{27}{mT}$, $B_\mathrm{max} = \SI{6}{T}$).}
    \newcolumntype{L}[1]{>{\raggedright\arraybackslash}p{#1}}
    \newcolumntype{C}[1]{>{\centering\arraybackslash}p{#1}}
    \newcolumntype{R}[1]{>{\raggedleft\arraybackslash}p{#1}}
    \newcommand{\ctab}{\centering\arraybackslash}
    \begin{tabular}{R{8mm}R{10mm}R{10mm}R{10mm}R{12mm}}
    \toprule
        \ctab$\alpha_\mathrm{p}$&\ctab$\hat{\theta}$    &\ctab$\sigma_\theta$   &\ctab$\hat{\theta}_\mathrm{meas}$
                                                                                                                &\ctab$\theta_\mathrm{start}^*$ \\
    \midrule
        \ang{0}                 &\ang{ 1.5}             &\ang{1.2}              &\SI{ 1.7(13)}{\degree}         &\ang{0.1}                      \\
        \ang{2}                 &\ang{11.5}             &\ang{1.3}              &\SI{ 5.7(34)}{\degree}         &\ang{0.9}                      \\
        \ang{4}                 &\ang{23.6}             &\ang{1.3}              &\SI{23.2(3)}{\degree}          &\ang{1.6}                      \\
        \ang{6}                 &\ang{36.8}             &\ang{1.5}              &\SI{38.2(2)}{\degree}          &\ang{2.4}                      \\
        \ang{8}                 &\ang{52.5}             &\ang{1.9}              &\SI{55.2(3)}{\degree}          &\ang{3.2}                      \\
        \ang{10}                &\ang{78.9}             &\ang{4.2}              &\SI{89.3(8)}{\degree}          &\ang{3.9}                      \\
        \ang{12}                & \multicolumn{3}{c}{--- magnetically reflected ---}                            &\ang{4.6}                      \\
    \bottomrule
    \end{tabular}
    \label{tab:sim_plate_distributions}
\end{table}

The measurements and simulations discussed in this work clearly show that the electron source achieves angular selectivity and can produce well-defined pitch angles with small angular spread.
Figure~\ref{fig:sim_plate_angles} shows the produced pitch angle in the spectrometer entrance magnet (solid red line) and the pitch angle at the end of the source chamber (dashed blue line) according to simulations. At $\alpha_\mathrm{max} = \ang{10.1}$, the pitch angle reaches \ang{90} and magnetic reflection occurs. Resulting from the finite angular spread, at $\alpha_\mathrm{p} = \ang{10}$ a fraction of the produced electrons is already reflected and cut off from the observed angular distribution.
The simulated reflection angle is in excellent agreement with the magnetic reflection measurement (section~\ref{measurements:reflection}), where $\alpha_\mathrm{max} = \SI{10.06(3)}{\degree}$ was observed for $\alpha_{az} = \ang{0}$.
The effective initial pitch angle shows a strictly linear dependency to the plate angle with a factor $k$ and a constant angular spread $\sigma_{\theta,\mathrm{start}} = \ang{0.1}$.

\section{Conclusion}
\label{conclusion}

An angular-selective electron source has been developed for the commissioning measurements of the KATRIN main spectrometer. In the first major measurement campaign at the KATRIN main spectrometer, several design improvements could be identified. After their implementation, the electron source was tested successfully at the monitor spectrometer in 2014. These preparation measurements demonstrated that the design requirements are completely fulfilled and that the electron source achieves all key features:
\begin{itemize}
    \item
        Angular selectivity: The source produces well-defined electron pitch angles in the spectrometer entrance magnet. Magnetic reflection occurs when the pitch angle exceeds \ang{90}, which was observed at a plate angle of \ang{10.1} in measurements. This value is in excellent agreement with the corresponding simulations, which also yield a reflection angle of \ang{10.1}.
    \item
        Small energy spread: Depending on the wavelength of the used UV light source, an energy spread between \SI{0.09(7)}{eV} at \SI{302}{nm} and \SI{0.031(5)}{eV} at \SI{266}{nm} was observed in transmission function measurements at $U_\mathrm{start} = \SI{-18.6}{kV}$. A measurement at low voltage $U_\mathrm{spec} \approx \SI{-200}{V}$ allows us to determine the energy spread with much higher precision because of the improved energy resolution of the spectrometer. While this feature cannot be applied at the monitor spectrometer, it is of great use for the commissioning of the main spectrometer.
    \item
        Small angular spread: At the monitor spectrometer setup with $B_\mathrm{max} = \SI{6}{T}$ and $B_\mathrm{start} = \SI{27}{mT}$, an angular spread of \ang{5} or less was observed in transmission function measurements at different plate angles. Simulations indicate that the angular spread is typically even smaller (about \ang{2}) for pitch angles $\theta \lesssim \ang{70}$.
    \item
        Electron rate: The electron source achieves a stable electron rate at the detector of \SI{1500}{cps} with the laser, and up to \SI{400}{cps} with the LEDs as light source. It is possible to regulate the rate by tuning the intensity of the UV photon system, \eg{} varying the pulse width of the LED pulser or by adjusting the laser diode current.
    \item
        Pulsed mode: The light sources were operated in pulsed mode during the monitor spectrometer measurements. The pulsed mode allows time-of-flight (ToF) measurements to characterize several properties of the MAC-E filter. The ToF mode plays an important role in the commissioning measurements of the main spectrometer.
\end{itemize}

The energy spread of the generated electrons depends on the work function of the photocathode and can be minimized by adjustung the UV wavelength to the properties of the utilized material. For our gold photocathode a work function of \SIerr{3.78}{0.03}{0.01}{eV} was found. Our value was determined \emph{in situ} and is considerably lower than the literature value for a clean gold surface that has been prepared under ultra-high vacuum conditions. This observation is explained by surface impurities from the continuous operation at $p \approx \SI{e-7}{mbar}$, where residual gas can be adsorbed onto the photocathode surface. Other effects, such as the unavoidable surface roughness and strong electric fields at the photocathode, can additionally lead to a reduction of the work function.
We used two different methods to directly determine the work function of the photocathode. A Fowler-type measurement, which investigates the wavelength-dependent electron yield, and a direct investigation of the energy distribution of the photo-electrons, which is derived from transmission function measurements. One advantage of the latter method is that the work function can be determined without requiring a dedicated wavelength scan. We demonstrated that this alternate method produces comparable results.

Particle-tracking simulations were performed with the Kassiopeia software, providing vital input for the analysis of the measurements, and allow us to get a precise understanding of the electron acceleration processes in the electron source. The simulation results are typically in good agreement with the measurements.
We showed that the energy distribution of the electrons in the spectrometer entrance magnet corresponds to to the initial energy distribution, while both distributions show the same width and shape in the simulations. It is thus possible to fully determine the electron energy spectrum by performing transmission function measurements with a MAC-E filter.
The angular distribution in the spectrometer magnet results from the non-adiabatic acceleration of the emitted electrons in the plate setup of the electron source and the subsequent adiabatic transport towards the spectrometer entrance. The electron beam is collimated by the strong electric acceleration field at the photocathode and reaches an effective angular spread of roughly \ang{0.1} when leaving the non-adiabatic acceleration region. According to simulations, an angular spread of \ang{2} (up to \ang{4} for $\theta \rightarrow \ang{90}$) is reached in the spectrometer magnet. This value is slightly lower than the measured angular spread of \ang{5}. The produced pitch angle and the angular spread in the magnet strongly depend on the magnetic fields at the setup. The differences between measurements and simulations can therefore be explained by undetected misalignments of the setup and entailing inaccuracies of the computed fields.

Our electron source allows us to investigate major characteristics of a MAC-E filter, such as the transmission properties and the effective energy resolution of the spectrometer. We studied key features of the electron source in measurements at the KATRIN monitor spectrometer and in a suite of accompanying simulations.
We fully characterized our electron source and demonstrated a reliable operation in a MAC-E filter setup. The electron source can be utilized as a vital tool for the commissioning of the KATRIN main spectrometer and in preparation of the upcoming neutrino mass measurements.

\begin{acknowledgement}
This work has been supported by the Helmholtz Association (HGF) and the Bundesministerium f{\"u}r Bildung und Forschung (BMBF) with project numbers 05A14PMA and 05A14VK2.
\end{acknowledgement}
\bibliography{references}

\begin{thebibliography}{10}
\providecommand{\url}[1]{{#1}}
\providecommand{\href}[2]{{#1}{#2}}
\providecommand{\urlprefix}{}
\providecommand{\doiprefix}{doi:}
\providecommand{\eprintprefix}{arXiv:}

\bibitem{DesignReport2005}
J.~Angrik, et~al., {KATRIN Design Report 2004, Wissenschaftliche Berichte FZKA
  7090}.
\newblock Tech. rep., Forschungszentrum Karlsruhe (2005).
\newblock
  \urlprefix\href{http://bibliothek.fzk.de/zb/berichte/FZKA7090.pdf}{http://bibliothek.fzk.de/zb/berichte/FZKA7090.pdf}

\bibitem{NuMass2013}
G.~Drexlin, et~al., Adv. High Energy Phys. \textbf{2013}, 293986 (2013),
  \doiprefix\href{http://dx.doi.org/10.1155/2013/293986}{10.1155/2013/293986}

\bibitem{NuMassMainz2005}
C.~Kraus, et~al., Eur. Phys. J. C \textbf{40}(4), 447 (2005),
  \doiprefix\href{http://dx.doi.org/10.1140/epjc/s2005-02139-7}{10.1140/epjc/s2005-02139-7}

\bibitem{NuMassTroitsk2011}
V.N. Aseev, et~al., Phys. Rev. D \textbf{84}, 112003 (2011),
  \doiprefix\href{http://dx.doi.org/10.1103/PhysRevD.84.112003}{10.1103/PhysRevD.84.112003}

\bibitem{NuMass2008}
E.W. Otten, C.~Weinheimer, Rep. Prog. Phys. \textbf{71}(8), 086201 (2008),
  \doiprefix\href{http://dx.doi.org/10.1088/0034-4885/71/8/086201}{10.1088/0034-4885/71/8/086201}

\bibitem{WGTS2015}
F.~Priester, et~al., Vacuum \textbf{116}, 42  (2015),
  \doiprefix\href{http://dx.doi.org/10.1016/j.vacuum.2015.02.030}{10.1016/j.vacuum.2015.02.030}

\bibitem{DPS2006}
X.~Luo, et~al., Vacuum \textbf{80}(8), 864  (2006),
  \doiprefix\href{http://dx.doi.org/10.1016/j.vacuum.2005.11.044}{10.1016/j.vacuum.2005.11.044}

\bibitem{DPS2012}
S.~Luki{\'c}, et~al., Vacuum \textbf{86}(8), 1126  (2012),
  \doiprefix\href{http://dx.doi.org/10.1016/j.vacuum.2011.10.017}{10.1016/j.vacuum.2011.10.017}

\bibitem{CPS2010}
W.~Gil, et~al., IEEE Trans. Appl. Supercond. \textbf{20}(3), 316 (2010),
  \doiprefix\href{http://dx.doi.org/10.1109/TASC.2009.2038581}{10.1109/TASC.2009.2038581}

\bibitem{MACE1992}
A.~Picard, et~al., Nucl. Instr. Meth. Phys. Res. B \textbf{63}(3), 345  (1992),
  \doiprefix\href{http://dx.doi.org/10.1016/0168-583X(92)95119-C}{10.1016/0168-583X(92)95119-C}

\bibitem{MACE1985}
V.M. Lobashev, P.E. Spivak, Nucl. Instr. Meth. Phys. Res. A \textbf{240}(2),
  305  (1985),
  \doiprefix\href{http://dx.doi.org/10.1016/0168-9002(85)90640-0}{10.1016/0168-9002(85)90640-0}

\bibitem{MACE1981}
G.~Beamson, et~al., J. Phys. A \textbf{14}(2), 256 (1981),
  \doiprefix\href{http://dx.doi.org/10.1088/0022-3735/14/2/526}{10.1088/0022-3735/14/2/526}

\bibitem{TritiumEndpoint2015}
E.G. Myers, et~al., Phys. Rev. Lett. \textbf{114}, 013003 (2015),
  \doiprefix\href{http://dx.doi.org/10.1103/PhysRevLett.114.013003}{10.1103/PhysRevLett.114.013003}

\bibitem{FPD2015}
J.F. Amsbaugh, et~al., Nucl. Instr. Meth. Phys. Res. A \textbf{778}, 40
  (2015),
  \doiprefix\href{http://dx.doi.org/10.1016/j.nima.2014.12.116}{10.1016/j.nima.2014.12.116}

\bibitem{WGTS2012}
M.~Babutzka, et~al., New J. Phys. \textbf{14}(10), 103046 (2012),
  \doiprefix\href{http://dx.doi.org/10.1088/1367-2630/14/10/103046}{10.1088/1367-2630/14/10/103046}

\bibitem{HVdivider2013}
S.~Bauer, et~al., J. Instrum. \textbf{8}(10), P10026 (2013),
  \doiprefix\href{http://dx.doi.org/10.1088/1748-0221/8/10/P10026}{10.1088/1748-0221/8/10/P10026}

\bibitem{HVdivider2009}
T.~Th{\"u}mmler, et~al., New J. Phys. \textbf{11}(10), 103007 (2009),
  \doiprefix\href{http://dx.doi.org/10.1088/1367-2630/11/10/103007}{10.1088/1367-2630/11/10/103007}

\bibitem{MonSpec2014}
M.~Erhard, et~al., J. Instrum. \textbf{9}(06), P06022 (2014),
  \doiprefix\href{http://dx.doi.org/10.1088/1748-0221/9/06/P06022}{10.1088/1748-0221/9/06/P06022}

\bibitem{ToF2013}
N.~Steinbrink, et~al., New J. Phys. \textbf{15}(11), 113020 (2013),
  \doiprefix\href{http://dx.doi.org/10.1088/1367-2630/15/11/113020}{10.1088/1367-2630/15/11/113020}

\bibitem{Egun2009}
K.~Valerius, et~al., New J. Phys. \textbf{11}(6), 063018 (2009),
  \doiprefix\href{http://dx.doi.org/10.1088/1367-2630/11/6/063018}{10.1088/1367-2630/11/6/063018}

\bibitem{Egun2011}
K.~Valerius, et~al., J. Instrum. \textbf{6}(01), P01002 (2011),
  \doiprefix\href{http://dx.doi.org/10.1088/1748-0221/6/01/P01002}{10.1088/1748-0221/6/01/P01002}

\bibitem{Egun2014}
M.~Beck, et~al., J. Instrum. \textbf{9}(11), P11020 (2014),
  \doiprefix\href{http://dx.doi.org/10.1088/1748-0221/9/11/P11020}{10.1088/1748-0221/9/11/P11020}

\bibitem{Kassiopeia2016}
D.~Furse, et~al.,   (2017).
\newblock \eprintprefix\href{https://arxiv.org/abs/1612.00262}{1612.00262}

\bibitem{Photoemission2012}
G.~Hechenblaikner, et~al., J. Appl. Phys. \textbf{111}(12), 124914 (2012),
  \doiprefix\href{http://dx.doi.org/10.1063/1.4730638}{10.1063/1.4730638}

\bibitem{Photoemission1931}
R.H. Fowler, Phys. Rev. \textbf{38}, 45 (1931),
  \doiprefix\href{http://dx.doi.org/10.1103/PhysRev.38.45}{10.1103/PhysRev.38.45}

\bibitem{Photoemission2002}
Z.~Pei, C.N. Berglund, Jpn. J. Appl. Phys. \textbf{41}(1A), L52 (2002),
  \doiprefix\href{http://dx.doi.org/10.1143/JJAP.41.L52}{10.1143/JJAP.41.L52}

\bibitem{groh:phd}
S.~Groh, {Modeling of the response function and measurement of transmission
  properties of the KATRIN experiment}.
\newblock Ph.D. thesis, {Karlsruher Institut f{\"u}r Technologie} (2015)

\bibitem{erhard:phd}
M.G. Erhard, {Influence of the magnetic field on the transmission
  characteristics and neutrino mass systematic of the KATRIN experiment}.
\newblock Ph.D. thesis, {Karlsruher Institut f{\"u}r Technologie} (2016)

\bibitem{behrens:phd}
J.D. Behrens, {Design and commissioning of a mono-energetic photoelectron
  source and active background reduction by magnetic pulse at the KATRIN
  spectrometers}.
\newblock Ph.D. thesis, {Westf{\"a}lische Wilhelms-Universit{\"a}t M{\"u}nster}
  (2016)

\bibitem{GeneralizedGaussian1997}
J.R.M. Hosking, J.R. Wallis, \emph{{Regional frequency analysis: an approach
  based on L-moments}} (Cambridge University Press, 1997), chap. A.8

\bibitem{emcee2013}
D.~Foreman-Mackey, et~al., Publ. Astron. Soc. Pac. \textbf{125}(925), 306
  (2013), \doiprefix\href{http://dx.doi.org/10.1086/670067}{10.1086/670067}

\bibitem{kraus:phd}
M.~Kraus, {Energy-scale systematics at the KATRIN main spectrometer}.
\newblock Ph.D. thesis, {Karlsruher Institut f{\"u}r Technologie} (2016)

\bibitem{barrett:phd}
J.P. Barrett, {A Spatially Resolved Study of the KATRIN Main Spectrometer Using
  a Novel Fast Multipole Method}.
\newblock Ph.D. thesis, {University of North Carolina at Chapel Hill} (2016)

\bibitem{wierman:phd}
K.~Wierman, {Charge Accumulation in the KATRIN Main Spectrometer}.
\newblock Ph.D. thesis, {University of North Carolina at Chapel Hill} (2016)

\bibitem{Photoemission1964a}
C.N. Berglund, W.E. Spicer, Phys. Rev. \textbf{136}, A1030 (1964),
  \doiprefix\href{http://dx.doi.org/10.1103/PhysRev.136.A1030}{10.1103/PhysRev.136.A1030}

\bibitem{MINUIT1975}
F.~James, M.~Roos, Comput. Phys. Commun. \textbf{10}(6), 343  (1975),
  \doiprefix\href{http://dx.doi.org/10.1016/0010-4655(75)90039-9}{10.1016/0010-4655(75)90039-9}

\bibitem{KelvinProbe1991}
M.~Nonnenmacher, et~al., Appl. Phys. Lett. \textbf{58}(25), 2921 (1991),
  \doiprefix\href{http://dx.doi.org/10.1063/1.105227}{10.1063/1.105227}

\bibitem{Photoemission1964b}
C.N. Berglund, W.E. Spicer, Phys. Rev. \textbf{136}, A1044 (1964),
  \doiprefix\href{http://dx.doi.org/10.1103/PhysRev.136.A1044}{10.1103/PhysRev.136.A1044}

\bibitem{Photoemission1995}
C.~Bandis, B.B. Pate, Phys. Rev. B \textbf{52}, 12056 (1995),
  \doiprefix\href{http://dx.doi.org/10.1103/PhysRevB.52.12056}{10.1103/PhysRevB.52.12056}

\bibitem{WorkFunction1970}
D.E. Eastman, Phys. Rev. B \textbf{2}, 1 (1970),
  \doiprefix\href{http://dx.doi.org/10.1103/PhysRevB.2.1}{10.1103/PhysRevB.2.1}

\bibitem{WorkFunction1995}
G.F. Saville, et~al., J. Vac. Sci. Technol. B \textbf{13}(6), 2184 (1995),
  \doiprefix\href{http://dx.doi.org/10.1116/1.588101}{10.1116/1.588101}

\bibitem{WorkFunctionEffects1973}
R.~D'Arcy, N.~Surplice, Surf. Sci. \textbf{34}(2), 193 (1973),
  \doiprefix\href{http://dx.doi.org/10.1016/0039-6028(73)90115-5}{10.1016/0039-6028(73)90115-5}

\bibitem{WorkFunctionEffects2005}
W.~Li, D.Y. Li, J. Chem. Phys \textbf{122}(6), 064708 (2005),
  \doiprefix\href{http://dx.doi.org/10.1063/1.1849135}{10.1063/1.1849135}

\bibitem{zacher:phd}
M.~Zacher, {High-field electrodes design and an angular-selective photoelectron
  source for the KATRIN spectrometers}.
\newblock Ph.D. thesis, {Westf{\"a}lische Wilhelms-Universit{\"a}t M{\"u}nster}
  (2014)

\bibitem{winzen:diploma}
D.~Winzen, {Development of an angular selective electron gun for the KATRIN
  main spectrometer}.
\newblock Diploma thesis, {Westf{\"a}lische Wilhelms-Universit{\"a}t
  M{\"u}nster} (2014)

\bibitem{RobinHood2012}
J.A. Formaggio, et~al., Prog. Electromagn. Res. B \textbf{39}, 1 (2012),
  \doiprefix\href{http://dx.doi.org/10.2528/PIERB11112106}{10.2528/PIERB11112106}

\bibitem{Elcd2011}
F.~Gl{\"u}ck, Prog. Electromagn. Res. B \textbf{32}, 319 (2011),
  \doiprefix\href{http://dx.doi.org/10.2528/PIERB11042106}{10.2528/PIERB11042106}

\bibitem{Magfield2011}
F.~Gl{\"u}ck, Prog. Electromagn. Res. B \textbf{32}, 351 (2011),
  \doiprefix\href{http://dx.doi.org/10.2528/PIERB11042108}{10.2528/PIERB11042108}

\bibitem{furse:phd}
D.L. Furse, {Techniques for Direct Neutrino Mass Measurement Utilizing Tritium
  $\beta$-Decay}.
\newblock Ph.D. thesis, {Massachusetts Institute of Technology} (2015)

\end{thebibliography}
\bibliographystyle{spphys-mod}
%
%
%
\end{document}